\documentclass[preprint2]{aastex}

\usepackage{amsmath}
\usepackage{graphics}

\newcommand \sigvlos {\sigma(v_{los})}
\newcommand \sigtand {\sigma(\tan{\delta})}
\newcommand \meanB   {\langle B\rangle}
\newcommand \rmsB    {\langle B^2 \rangle}
\newcommand \Brms    {B_{rms}}
\newcommand \Bcfmod  {B_{CF}^{mod}}

\begin{document}

\title{Magnetic Field Diagnostics Based on Far-Infrared Polarimetry:
       Tests Using Numerical Simulations}

\author{Fabian Heitsch\altaffilmark{1}}
\author{Ellen G. Zweibel\altaffilmark{2}}
\author{Mordecai-Mark Mac Low\altaffilmark{3}}
\author{Pakshing Li\altaffilmark{4}}
\author{Michael L. Norman\altaffilmark{4}}
\altaffiltext{1}{Max-Planck-Institut f\"ur Astronomie, K\"onigstuhl 17, D-69117
Heidelberg, Germany; E-mail: heitsch@mpia-hd.mpg.de}
\altaffiltext{2}{JILA, University of Colorado, Campus Box 440, 
Boulder, CO 80309, USA; E-mail: zweibel@solarz.colorado.edu}
\altaffiltext{3}{Department of Astrophysics, American Museum of
Natural History, Central Park West at 79th Street, New York, New York
10024-5192, USA; E-mail: mordecai@amnh.org}
\altaffiltext{4}{LCA, National Center for Supercomputing Applications, University
of Illinois at Urbana-Champaign, Urbana-Champaign, IL 61801, USA}

\begin{abstract}
The dynamical state of star-forming molecular clouds cannot be understood
without determining the structure and strength of their magnetic fields.
Measurements of polarized far-infrared radiation from thermally aligned dust
grains are used to map the orientation of the field and estimate its strength,
but the accuracy of the results has remained in doubt. In order to assess
the reliability of this method, we apply it to simulated far-infrared
polarization maps derived from three-dimensional simulations of supersonic
magnetohydrodynamical turbulence, and compare the estimated values to
the known magnetic field strengths in the simulations. We investigate the effects of limited 
telescope resolution and self-gravity on the structure of the maps. Limited observational
resolution affects the  field structure such that small scale variations can be completely
suppressed, thus giving the impression of a very homogeneous field.
The Chandrasekhar-Fermi method of estimating the mean magnetic field in a turbulent medium 
is tested, and we suggest an extension to measure the rms field. Both methods yield results 
within a factor of $2$ for field strengths typical of molecular clouds, with the modified 
version returning more reliable estimates for slightly weaker fields. However, neither method
alone works well for very weak fields, missing them by a factor of up to $150$. Taking the 
geometric mean of both methods estimates even the weakest fields accurately within a factor of $2.5$.
Limited telescope resolution leads to a systematic 
overestimation of the field strengths for all methods. 
We discuss the effects responsible for this overestimation 
and show how to extract information on the underlying (turbulent) power spectrum.
\end{abstract}


\clearpage

%%%%%%%%%%%%%%%%%%%%%%%%%%%%%%%%%%%%%%%%%%%%
%%%%%%%%%%%%%%%%%%%%%%%%%%%%%%%%%%%%%%%%%%%%
%
% section{Motivation}
%
%%%%%%%%%%%%%%%%%%%%%%%%%%%%%%%%%%%%%%%%%%%%
%%%%%%%%%%%%%%%%%%%%%%%%%%%%%%%%%%%%%%%%%%%%
\section{Motivation\label{sec:motivation}}

The significance of magnetic fields in the dynamics of molecular clouds, and in 
star formation itself, is still uncertain. 

Theory and modelling have demonstrated that the strength
of magnetic fields can have a profound influence on the processes which lead
to the formation of stars. There is a critical ratio of magnetic flux to mass
above which magnetic fields can prevent gravitational collapse (magnetically
subcritical) and below which they cannot (magnetically supercritical; 
e.g. Mouschovias \& Spitzer 1976). Redistribution of magnetic flux by
ambipolar diffusion in a magnetically subcritical region can lead to 
loss of support against self-gravitation, and eventually to 
low mass star formation (see Shu, Adams, \& Lizano 1987, Myers \& Goodman 
(1988), Porro \& Silvestro (1993), Ciolek \& Mouschovias 1993, 1994, 1995,
Safier, McKee, \& Stahler 1997, Ciolek \& K\"onigl 1998). However,
magnetically subcritical regions do not appear to reproduce observations of 
molecular cloud cores (Nakano 1998), nor can this scenario explain the large
fraction of cloud cores containing protostellar objects 
(Ward-Thompson, Motte, \& Andr\'{e} 1999). It is possible that magnetically
supercritical regions can be supported by turbulence (Bonazzola et al. 1987, McKee
\& Zweibel 1995), which, to support a highly supercritical region, must be
quite nonlinear (Myers \& Zweibel 2001). Recent 3D
simulations (Heitsch, Mac Low, \& Klessen 2001, hereafter HMK, Ostriker, 
Stone, \& Gammie 2001) demonstrate that strong turbulence can provide
large-scale support against collapse, though it cannot prevent collapse
on small scales. The degree to which clouds are supported at all, of course,
depends on their lifetimes and on the star formation rate 
(Ballesteros-Paredes, Hartmann, \& Vazquez-Semadeni 1999, Elmegreen 2000).
 
Magnetic fields affect other dynamical processes in clouds besides 
gravitational collapse.  In a strong magnetic field, weakly compressible
turbulence is anisotropic (Sridhar \& Goldreich 1994; Goldreich \& Sridhar 1995, 1997) 
and energy dissipation is relatively more in waves and less in shocks 
(Smith, Mac Low, \& Zuev 2000), although the
overall rate of energy dissipation is not strongly dependent on the
fieldstrength (Mac Low 1999). Magnetic fields may also play a role in
collimating molecular outflows, and in transferring their momentum to the
ambient medium. 

The actual strength of magnetic fields in molecular clouds will ultimately
determine which theoretical picture is correct, so observations of field
strengths are crucial. The integrated line of sight
component of the field, weighted by the density of the tracer species, can be
measured through the Zeeman effect. However, Zeeman mapping is time consuming
and requires high sensitivity and the presence of particular tracers (most
commonly OH). Moreover, this technique does not probe the field in the plane 
of the sky. For all of these reasons, it is desirable to have a complementary 
method of mapping the field.
 
Following the first detection by Cudlip et al. (1982), it has been shown that
the polarization of the far infrared thermal radiation emitted by magnetically
aligned dust grains can be used to map the orientation of the magnetic field
on the plane of the sky (see Hildebrand et al 2000 for methodology). 
This has been done, at various wavelengths, for
a number of nearby clouds (Gonatas et al. 1990, Jarrett et
al 1994, Dotson 1996, Rao et al. 1998, Schleuning 1998, Glenn, Walker, \&
Young 1999, Dotson et al. 2000, Schleuning et al. 2000, Vall\'{e}e, Bastien, \& Greaves 2000,
Ward-Thompson et al 2000) and near the 
Galactic Center (Werner et al. 1988, Hildebrand et al. 1990, 1993,
Dowell 1997, Novak et al. 1997, 2000). Far infrared polarization maps display the
morphology of the field relative to other structures in the cloud, and can
also be used to estimate the strength of the mean field according to a
dynamical method originally proposed by Chandrasekhar \& Fermi (1953, hereafter
CF). Applications of the CF-method generally suggest fieldstrengths in the
milligauss range or above. Such fieldstrengths are larger than what is
typically measured by the Zeeman effect (e.g. Glenn et al. 1999, Lai et al. 2000).

The CF-method is subject to errors arising from line of sight and angular
averaging, and rests on the assumptions of equipartition between turbulent
kinetic and magnetic energy, and isotropy of fluid motions. 
Spatial averaging, by smoothing the maps, can
give misleading impressions about the magnetic field morphology. However,
since the information carried in the maps is unique, it is difficult to
test the magnitude of these effects with astronomical observations.

Numerical simulations of turbulent, magnetized molecular clouds
offer the means to calibrate the accuracy of polarization
maps and develop new techniques to analyze them. 
The actual strength and structure of the field are known 
at all gridpoints and at selected times, as are the gas velocity and density.
It is possible to create synthetic polarization maps, analyze them as though
they were astronomical data, and check the accuracy of the results. That is the
subject of this paper. Ostriker,
Stone, \& Gammie (2001) have used a set of simulations to 
calibrate the CF-method in this manner.

In section \ref{sec:models-methods}, we describe the numerical models and the 
method by which we generate polarization maps. Most of the results relevant 
to polarization maps are in section \ref{sec:results}, in which we discuss the morphology 
shown in the maps, implement the CF-method, devise an alternative to it, and show what 
can be learned about the spectrum of magnetic field fluctuations. 
Section \ref{sec:conclusions} is a summary and discussion.

%%%%%%%%%%%%%%%%%%%%%%%%%%%%%%%%%%%%%%%%%%%%
%%%%%%%%%%%%%%%%%%%%%%%%%%%%%%%%%%%%%%%%%%%%
%
% section{Models and Methods}
%
%%%%%%%%%%%%%%%%%%%%%%%%%%%%%%%%%%%%%%%%%%%%
%%%%%%%%%%%%%%%%%%%%%%%%%%%%%%%%%%%%%%%%%%%%
\section{Models and Methods\label{sec:models-methods}}

%%%%%%%%%%%%%%%%%%%%%%%%%%%%%%%%%%%%%%%%%%%%
%%%%%%%%%%%%%%%%%%%%%%%%%%%%%%%%%%%%%%%%%%%%
% subsection{Models of 3D-MHD-turbulence}
%%%%%%%%%%%%%%%%%%%%%%%%%%%%%%%%%%%%%%%%%%%%
%%%%%%%%%%%%%%%%%%%%%%%%%%%%%%%%%%%%%%%%%%%%
\subsection{Models of 3D-MHD-turbulence\label{subsec:models-models}}

We base our investigation on full 3D models of driven MHD-turbulence in a
cube with periodic boundary conditions, simulating a portion of the interior of
a molecular cloud. We chose an isothermal equation of state, because
the cooling times are much shorter than the dynamical times at the high densities
typical of molecular clouds. We performed the simulations at $128^3$ and $256^3$
grid zones resolution using ZEUS-3D, a well-tested Eulerian finite-difference 
code (Stone \& Norman 1992a, 1992b, Clarke 1994) with second-order advection and a 
von Neumann artificial viscosity to capture shocks. The MHD induction equation
is followed using the method of consistent transport along characteristics
(Hawley \& Stone 1995). We employed the massively-parallel version of the code, 
ZEUS-MP (Fiedler 1998, Norman 2000), to produce a data set at resolution $512^3$. 

We used the uniform driving 
mechanism described in Mac Low (1999). At each timestep, a fixed pattern
of velocity perturbations is added, with the amplitude adjusted such that 
the energy input rate is kept constant. The driving results in an {\em rms} 
Mach number of ${\cal M} = 5$ for models of series ${\cal G}$ and of 
${\cal M} = 10$ for such of series ${\cal E}$ (see Table \ref{tab:models}). 
The mechanism for generating the perturbation field allows us to select fixed 
spatial ranges. All models presented in this work employ driving at 
wavenumbers k=$1-2$ waves per box length $L$.

Our measurements begin at system time $t$ = 0.0, when the model has reached an
equilibrium state between the energy dissipation rate due to shock interaction
and numerical diffusion, and the driving energy input rate.
Self-gravity is implemented via an FFT-Poisson solver for Cartesian coordinates
(Burkert \& Bodenheimer 1993). It is activated as soon as the model reaches
the dissipation equilibrium at $t$ = 0.0.

The isothermal equation of state renders the system scale free. In code units,
the length of the box is $L=2$ and the mass is $M=1$. We dimensionalize the
results by choosing the Jeans length $\lambda_J$, Jeans mass $M_J$, and free fall
time $t_{ff}$ (see Klessen, Heitsch, \& Mac Low 2000 for discussion of the
scaling). The number of Jeans masses and Jeans lengths in
each run are given in Table \ref{tab:models}.

All the models are initialized with a uniform magnetic field stretching across
the box along the $z$-direction. The field becomes distorted over time, but the 
magnetic flux $\Phi$ through the boundaries should be constant according to
the equations of ideal MHD, and is typically preserved by the code to a
relative accuracy of 10$^{-3}$ - 10$^{-4}$.

As in HMK, we scale the initial magnetic fieldstrength
by the plasma beta $\beta\equiv 8\pi P/B^2$. The parameter $\beta$
is directly related to the critical magnetic flux required to prevent global
gravitational collapse (see McKee et al 1993)
\begin{equation}
  \Phi_c = \frac{G^{1/2}M}{c_{\Phi}},
  \label{equ:magstatsup}
\end{equation}
where the constant $c_\Phi$ is found to be 0.13 for a uniformly magnetized
sphere (Mouschovias \& Spitzer 1976) and 0.16 for a uniformly magnetized
sheet (Nakano \& Nakamura 1978). Here we choose the latter value. We find
\begin{equation}
  \beta = 8\left(\frac{\Phi_c}{\Phi}\frac{\lambda_J}{L}c_{\Phi}\right)^2
        = 8\left(\frac{M}{M_c}\frac{\lambda_J}{L}c_{\Phi}\right)^2,
\end{equation}
where $M/M_c=\Phi_c/\Phi$ is the ratio of mass to critical mass as in HMK.
The models discussed in this paper span the range $\beta = 0.05 - 4.04$ and
$M/M_c = 0.88 - 8.3$.  The Alfv\'{e}n Mach number ${\cal M}_A$
is related to $\beta$ and the sonic Mach number ${\cal M}$ by ${\cal M}_A =
{\cal M}(\beta/2)^{1/2}$. Thus, we consider both sub-Alfv\'{e}nic and 
super-Alfv\'{e}nic models.

A detailed discussion of the model sets is given by HMK.
Here we describe their physical properties just briefly.
The supersonic turbulence supports the gravitationally unstable region against global
collapse, however, it cannot prevent local collapse, even in the presence of magnetic
fields too weak to provide magnetostatic support. Collapsing regions evolve
from shock-induced filaments and cores. The models reach density contrasts of 
$2-3$ orders of magnitude above and below the mean density. The gravitationally
bound regions can be interpreted as the initial stages of a protostellar core which 
may subsequently evolve into stars. 
Once a core begins to collapse, we cannot resolve it well enough 
to follow its evolution further (Truelove et al. 1997, HMK). 

The isotropy of a model depends largely on its field strength.
For the weak-field-model ${\cal E}h1a$, the resulting distribution of magnetic 
energy is fairly isotropic, whereas for the models with stronger fields 
(e.g. ${\cal E}h1d$), the field imprints its initial direction onto the gas 
flows. The flows, in turn, act on the field. Two quantities of interest for
the polarization studies are the ratio of the mean field energy to the total
magnetic energy, $\langle B\rangle^2/\langle B^2\rangle$, and the dispersion
in the angle $\delta$ between the local magnetic field direction and the mean
direction, $\sigma^2(\delta)$
(the two are not equivalent; for example, the energy
in the field could be increased by
collecting it into strong unidirectional filaments with no dispersion in
angle). Both quantities are given in Table \ref{tab:turbcharacter}.

Equipartition between turbulent kinetic and turbulent magnetic energy is often 
assumed in astrophysics; it has been shown to hold rigorously only in 
certain cases, such 
as weak Alfv\'{e}n wave turbulence (Zweibel \& McKee 1995) and the 
incompressible Alfv\'{e}nic turbulence modelled by Goldreich \& Sridhar
(1997). Observations
do not yet yield clear evidence for or against equipartition, although
data collected by Crutcher (1999) tend to speak against it. 
The models discussed here are not in exact equipartition; the ratio 
$\xi=E_{mag}^{turb}/E_{kin}$ of turbulent magnetic to turbulent kinetic energy 
for each model is listed in Table \ref{tab:turbcharacter}, and is typically
a few tenths. It is unclear whether these departures from equipartition occur 
for physical reasons related to the nature of nonlinear, compressible MHD
turbulence or occur because of numerical diffusivity or the nature of the
forcing. We take the deviation from equipartition into account in 
\S~\ref{subsec:res-cf} to correctly interpret our polarization maps.

%%%%%%%%%%%%%%%%%%%%%%%%%%%%%%%%%%%%%%%%%%%%
%%%%%%%%%%%%%%%%%%%%%%%%%%%%%%%%%%%%%%%%%%%%
% subsection{Generating Polarization Maps}
%%%%%%%%%%%%%%%%%%%%%%%%%%%%%%%%%%%%%%%%%%%%
%%%%%%%%%%%%%%%%%%%%%%%%%%%%%%%%%%%%%%%%%%%%
\subsection{Generating Polarization Maps\label{subsec:models-maps}}

The model cubes include density, magnetic fields and velocity fields. 
These data allow us to derive the Stokes parameters $Q$ and $U$, and 
the polarized intensity $P$ according to Zweibel (1996)
\begin{equation}
  P = Q + iU \propto \int f(y) \frac{(B_x + i B_z)^2}{B_x^2 + B_z^2} dy
  \label{equ:stokesparm}
\end{equation}
where $B_x$ and $B_z$ give the field vectors in the plane of sky perpendicular
to the line of sight and are taken directly from the simulations. We integrate 
along the line of sight in $y$-direction.
 
The function $f(y)$ is a weighting function which accounts for the density,
emissivity, and polarizing properties of the dust grains. In this paper, we
take it to be the gas density normalized by the mean density. 
The polarized
intensity is $|P| = \sqrt{Q^2 + U^2}$, and the polarization angle is
\begin{equation}
  \phi = \frac{1}{2}\arctan{\frac{Q}{U}}.
  \label{equ:polarangle}
\end{equation}

Equation (\ref{equ:stokesparm}) is an approximate solution of the full radiative
transfer equation for the Stokes parameters (Martin 1974, Lee \& Draine 1985),
valid for small polarization and low optical depth. At far infrared 
wavelengths, and at typical column densities for molecular clouds, the
medium can safely be assumed to be optically thin (Hildebrand et al. 2000).

Polarization maps are made at finite resolution. We 
simulate the telescope beam by 
applying a Gaussian filter of the form
\begin{equation}
  F(x,x') = \frac{1}{\sqrt{2\pi}w} \exp\left(-\frac{1}{2}\left(\frac{x-x'}{w}\right)^2\right)
  \label{equ:gaussfilter}
\end{equation}
to the complex polarization $P$ given in equation (\ref{equ:stokesparm}). 
The width $w$ of the smoothing filter should not exceed one eighth of the
box length $L$. For $w > L/8$, the tails of the filter -- exceeding the 
map's area -- would contribute to such an extent that neglecting 
them would yield a too small average. For each model, we generated 
polarization maps with a set of smoothing widths $0 \leq w \leq L/8$.

%%%%%%%%%%%%%%%%%%%%%%%%%%%%%%%%%%%%%%%%%%%%
%%%%%%%%%%%%%%%%%%%%%%%%%%%%%%%%%%%%%%%%%%%%
%
% section{Results}
% 
%%%%%%%%%%%%%%%%%%%%%%%%%%%%%%%%%%%%%%%%%%%%
%%%%%%%%%%%%%%%%%%%%%%%%%%%%%%%%%%%%%%%%%%%%
\section{Results\label{sec:results}}

The 2D polarization maps serve
a threefold purpose: We discuss how their structure depends on 
self-gravity and limited observational resolution 
(\S~\ref{subsec:res-struct}), and we show that there is no preferred alignment 
between filamentary structures in our simulations and the magnetic fields 
(\S~\ref{subsec:res-filament}). Finally, in \S~\ref{subsec:res-cf} we 
discuss the Chandrasekhar-Fermi method of determining the mean field
and an extension, which determines the rms field,
both of which we test and calibrate.

%%%%%%%%%%%%%%%%%%%%%%%%%%%%%%%%%%%%%%%%%%%%
%%%%%%%%%%%%%%%%%%%%%%%%%%%%%%%%%%%%%%%%%%%%
% subsection{Structure in Polarization Maps
%%%%%%%%%%%%%%%%%%%%%%%%%%%%%%%%%%%%%%%%%%%%
%%%%%%%%%%%%%%%%%%%%%%%%%%%%%%%%%%%%%%%%%%%%
\subsection{Structure in Polarization Maps\label{subsec:res-struct}}

\begin{figure*}
\begin{center}
\includegraphics[width=16.0cm, bb=70 220 550 700]{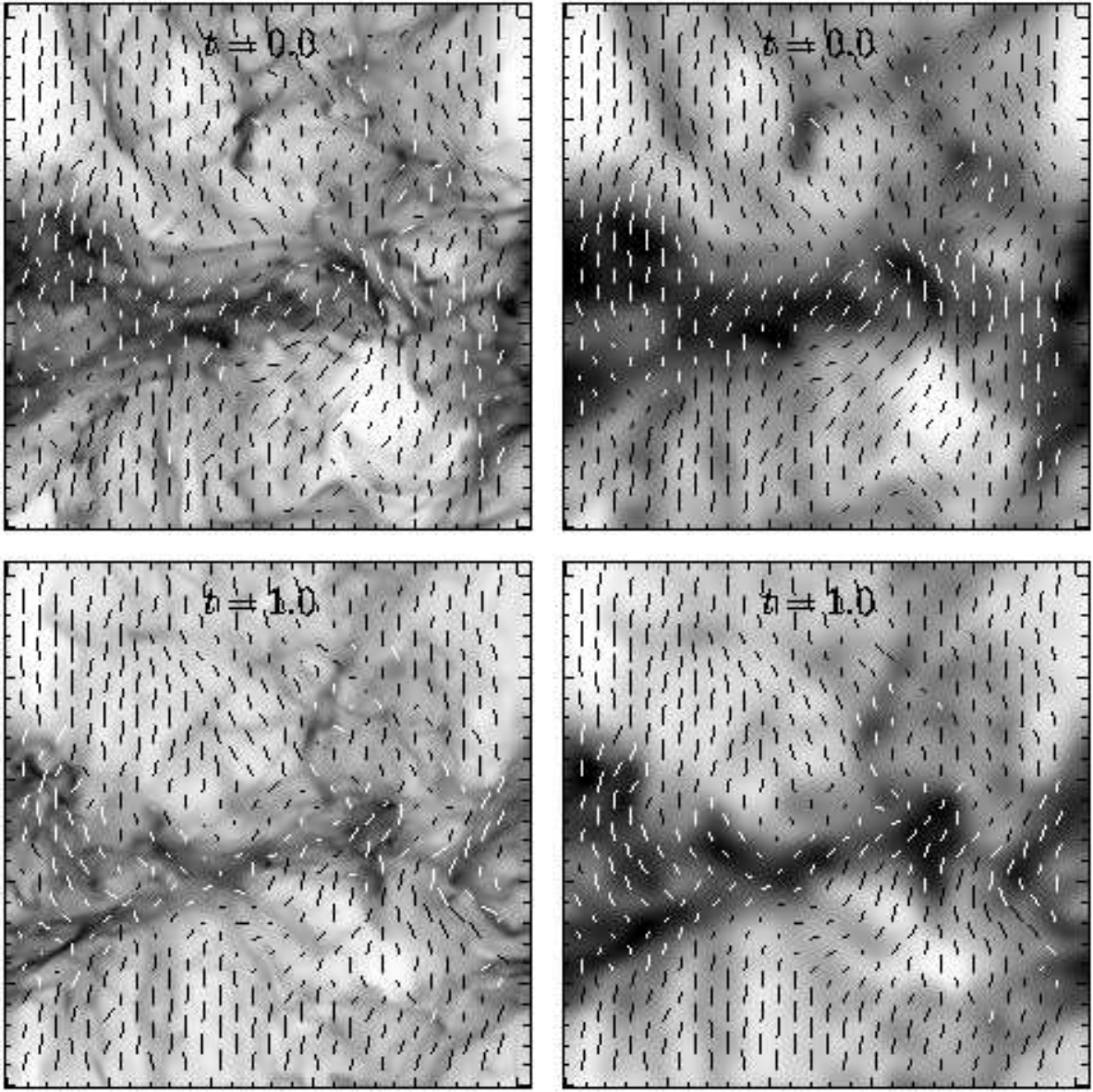}
\end{center}
\caption{\label{fig:polmap-maps}
{\em Left column:}
Surface density (corresponding to continuum density) 
of the full computational domain, overplotted with
polarization vectors as determined by equation~(\ref{equ:stokesparm})
for the self-gravitating model ${\cal E}h1d$. The magnetic field is
supercritical with $M/M_{c}= 1.8$. 
At $t=0.0$ (upper panel), gravity has been switched on, which shows 
its effect in the lower panel, at $t=1.0$, a free-fall time later.
The initially uniform field is vertically orientated (as well as
for all other maps shown).
{\em Right column:}
Surface density and polarization vectors as in left column,
but smoothed with a Gaussian filter of 
$w = 4$ pixel. The field gets more ordered and the previously well
discernible shock structures are blurred or even come out as clumpy,
filamentary structures.}
\end{figure*}

The polarization maps showing column density and polarization vectors
(Fig. \ref{fig:polmap-maps} left column, model ${\cal E}h1d$) are highly structured. 
Shock fronts moving through the gas initiate formation of filaments and knots.
After one free-fall time (lower left panel in Fig. \ref{fig:polmap-maps}), the filaments
fragment and concentrate due to self-gravity. Qualitatively, no influence of self-gravity 
on the large-scale structure of the field is discernible, although there is 
some effect on the smallest scales.

Smoothing these maps with a Gaussian filter of $4$ pixels width
(Fig. \ref{fig:polmap-maps} right column, ${\cal E}h1d$) leads to a clumpier 
appearance of the previously well-defined filaments. Any substructure 
in these is lost. Single shock structures are smeared out.
As expected, the field appears more uniform, thus in our case indicating 
more and more its initial orientation. Small scale variations in the field
indicating turbulence and defining the turbulent cascade are lost due to the 
smoothing. 

\begin{figure*}
\begin{center}
\includegraphics[width=12.0cm,angle=90]{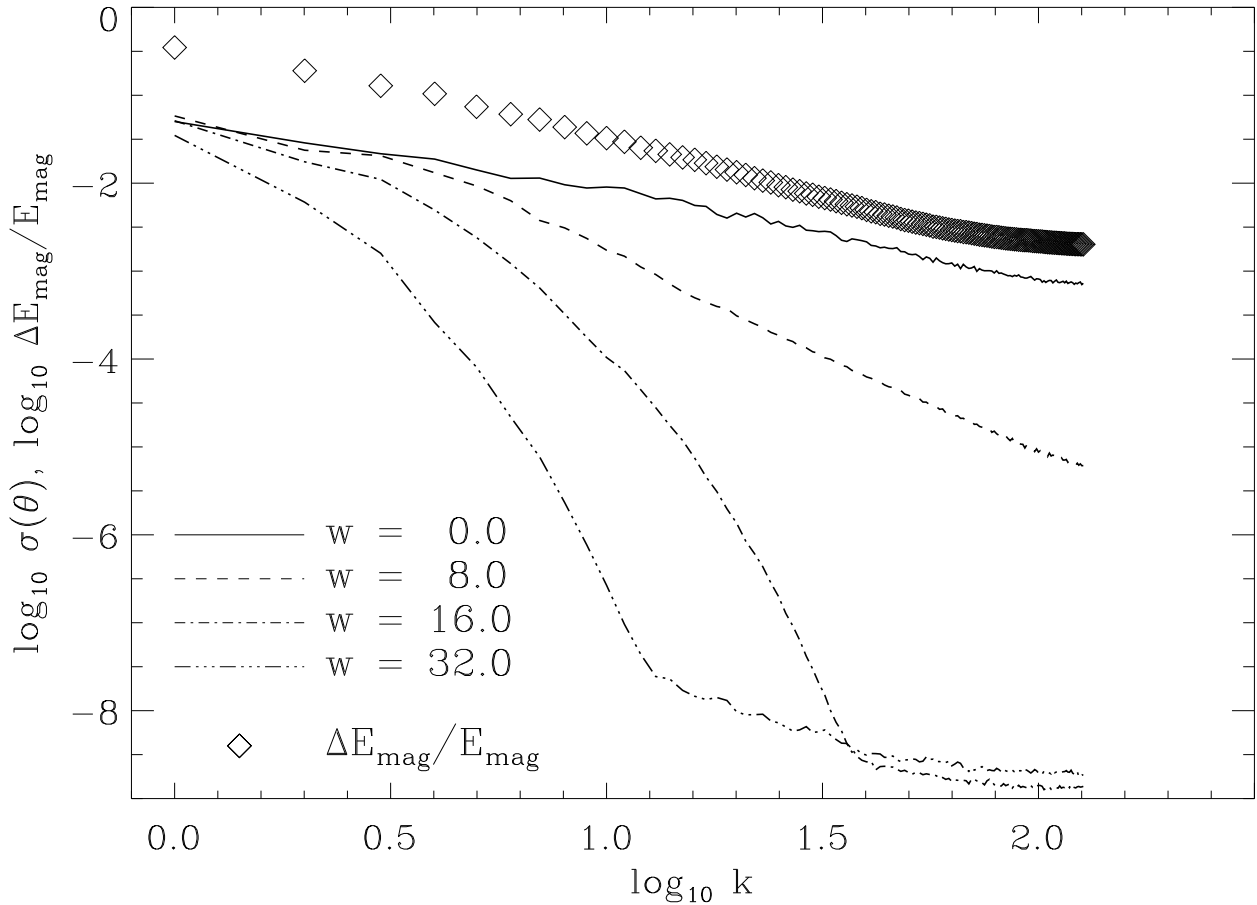}
\end{center}
\caption{\label{fig:smoothphi}
%fig-smoothphi.ps
Power spectrum of polarization angles with respect to mean magnetic field
of model ${\cal E}h1d$ and ratio $\Delta E_{mag}/E_{mag}$, corresponding
to $\langle B_x^2\rangle/\langle B_y\rangle^2$, the perturbed against
mean field (diamonds) against wavenumber $k$. 
The energy ratio is shifted horizontally by
$\Delta \log(\Delta E_{mag}/E_{mag})=0.5$ for clarity.
$w$ denotes the smoothing width, $0 \leq w \leq L/8$, 
with $L=256$ the box length. At $w=L/8$, $96$\% of the power 
in polarization angles is lost.}
\end{figure*}

Figure \ref{fig:smoothphi} quantifies this effect in model ${\cal E}h1d$. 
It shows the power spectrum of the angle between the local and the mean magnetic field.
This is equivalent to the line of sight averaged spectrum of the magnetic
field fluctuation amplitude (recall that the total power in fluctuations is
given in Table \ref{tab:turbcharacter}).
Increasing the smoothing width from $w=0$ to $w=L/8$ results in a power loss
of $96$\%. Structures at a wave number of $k=8$ are suppressed by 
more than two orders of magnitude. 
The diamonds in Figure \ref{fig:smoothphi} indicate the power spectrum of
perturbed against mean field energy $E_{mag}^{turb}/E_{mag}$,
where $E_{mag}^{turb}$ is the magnetic energy corresponding to the 
perturbed field components perpendicular to the mean field.
We use this as a measure of the true disorder in
the magnetic field and as a gross check whether the polarization angles mirror 
the behaviour of the energies. Some power loss between the true angular dispersion and
even the unsmoothed fluctuation amplitude is inevitable, because the latter
involves line of sight averaging while the former is a true 3D quantity.

\begin{figure*}
\begin{center}
\includegraphics[width=12.0cm]{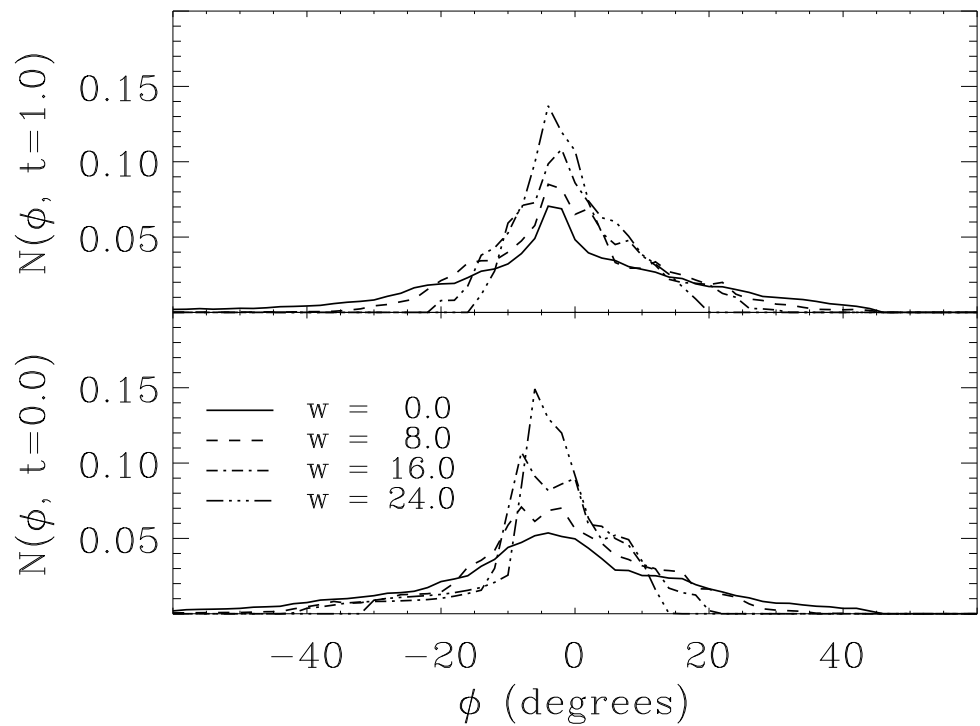}
\end{center}
\caption{\label{fig:ang-dist}
%fig-gravsig.ps
Distribution of polarization angles for two times and various
smoothing beam widths $w$ of model ${\cal E}h1d$. The bottom panel starts
with $t=0.0$, when gravity is switched on. The upper panel shows the angle
distribution one free-fall time later.
With increasing $w$, the distributions get narrower,
whereas self-gravity does not affect the distributions significantly except
for the smallest scales ($w=0$).}
\end{figure*}

Inspection of Figure (\ref{fig:polmap-maps}) 
shows no dramatic effects caused by self gravity after one free-fall time. 
Figure (\ref{fig:ang-dist}) shows that 
the width of the distribution of polarization angles does not change significantly 
under the effect of self-gravitation, but does certainly decrease with increasing beam width,
as expected. The emerging asymmetry results from the reduced statistics due to the 
increasing beam width. Although every pixel contributes to the histogram, the number of
independent measurements decreases with increasing beam width.

%%%%%%%%%%%%%%%%%%%%%%%%%%%%%%%%%%%%%%%%%%%%
%%%%%%%%%%%%%%%%%%%%%%%%%%%%%%%%%%%%%%%%%%%%
% subsection{Shock-induced Filaments}
%%%%%%%%%%%%%%%%%%%%%%%%%%%%%%%%%%%%%%%%%%%%
%%%%%%%%%%%%%%%%%%%%%%%%%%%%%%%%%%%%%%%%%%%%
\subsection{Shock-induced Filaments\label{subsec:res-filament}}

There is an ongoing debate on the mechanisms generating the 
observed filamentary structures of molecular clouds (e.g. Loren 1989,
Johnstone \& Bally 1999, Matthews \& Wilson 2000). 
Shocks come to mind as a natural explanation, either compressing the 
material (as in the simulations presented) or generating downstream
flows resulting in the filaments (Loren 1989). The alignment of 
the magnetic field with the filaments has been used as a test for
the validity of different filament models. The observations seem to 
favour no definite alignment. Maps of OMC-III (Matthews \& Wilson 2000)
show a perpendicular alignment, thus perhaps supporting the model of
Fiege \& Pudritz (2000a, 2000b), in which they propose helical fields
to confine filamentary structures. Rizzo et al. (1998) found parallel
and perpendicular alignment in their maps of background starlight 
polarization for Lupus 1 and Lupus 4, whereas Rao et al. (1998) 
find varying alignments for Orion-KL. Further examples of varying
field alignments can be found in Dotson (1996) and Ward-Thompson et al. (2000).

\begin{figure*}
\begin{center}
\includegraphics[bb=80 80 520 520]{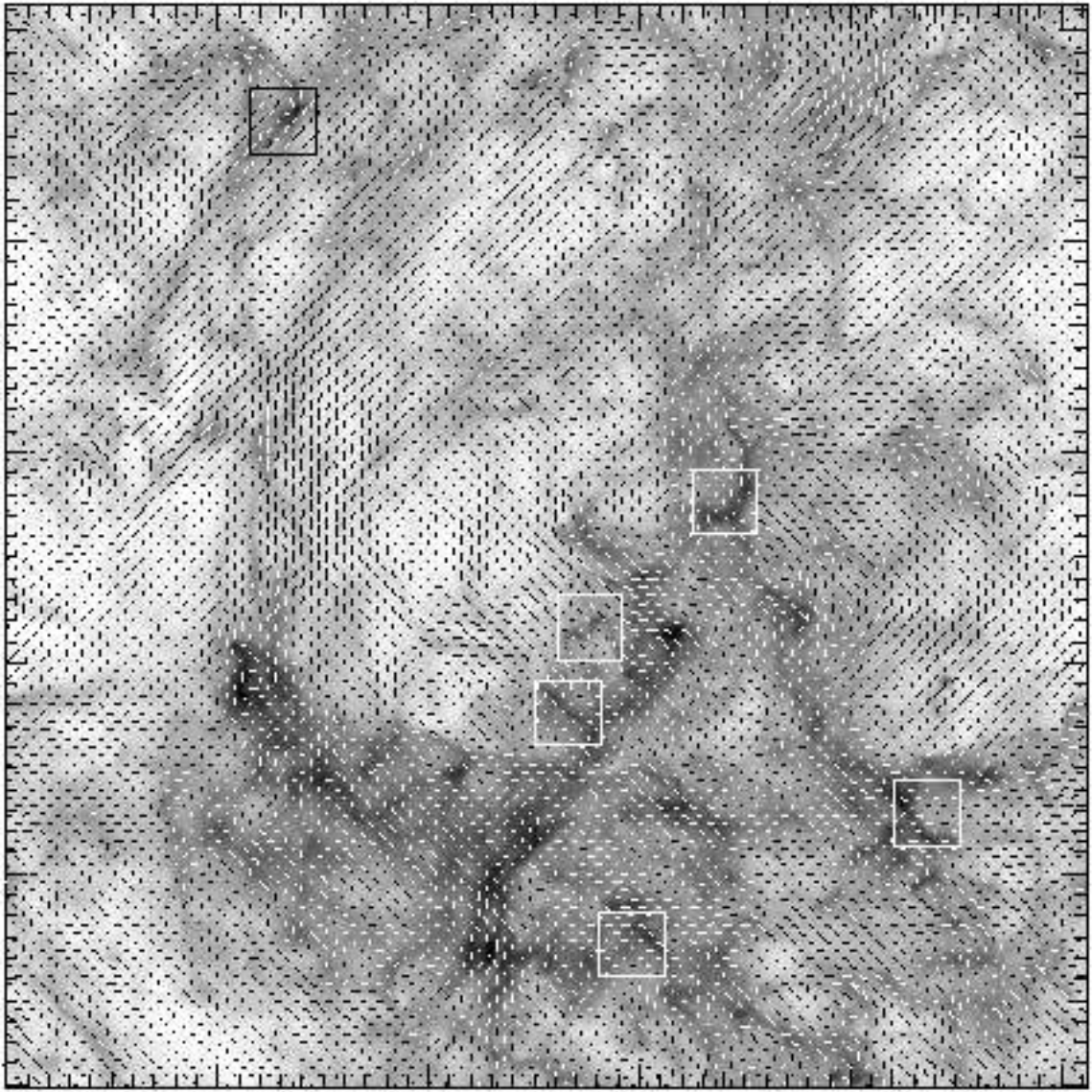}
\end{center}
\caption{\label{fig:highresmap}
%fig-weakfield.ps
Surface density of the full computational domain of $512^2$ cells, 
overplotted with magnetic
polarization vectors as determined by equation~(\ref{equ:stokesparm})
for the not yet self-gravitating model ${\cal E}r1a$ at $t=0.0$. The 
magnetic field is supercritical by a factor of $8.3$. The white boxes 
mark the filament locations as shown in Figure \ref{fig:highresfil}.
Every eighth polarization vector is shown.}
\end{figure*}

\begin{figure*}
\begin{center}
\includegraphics[height=21.0cm]{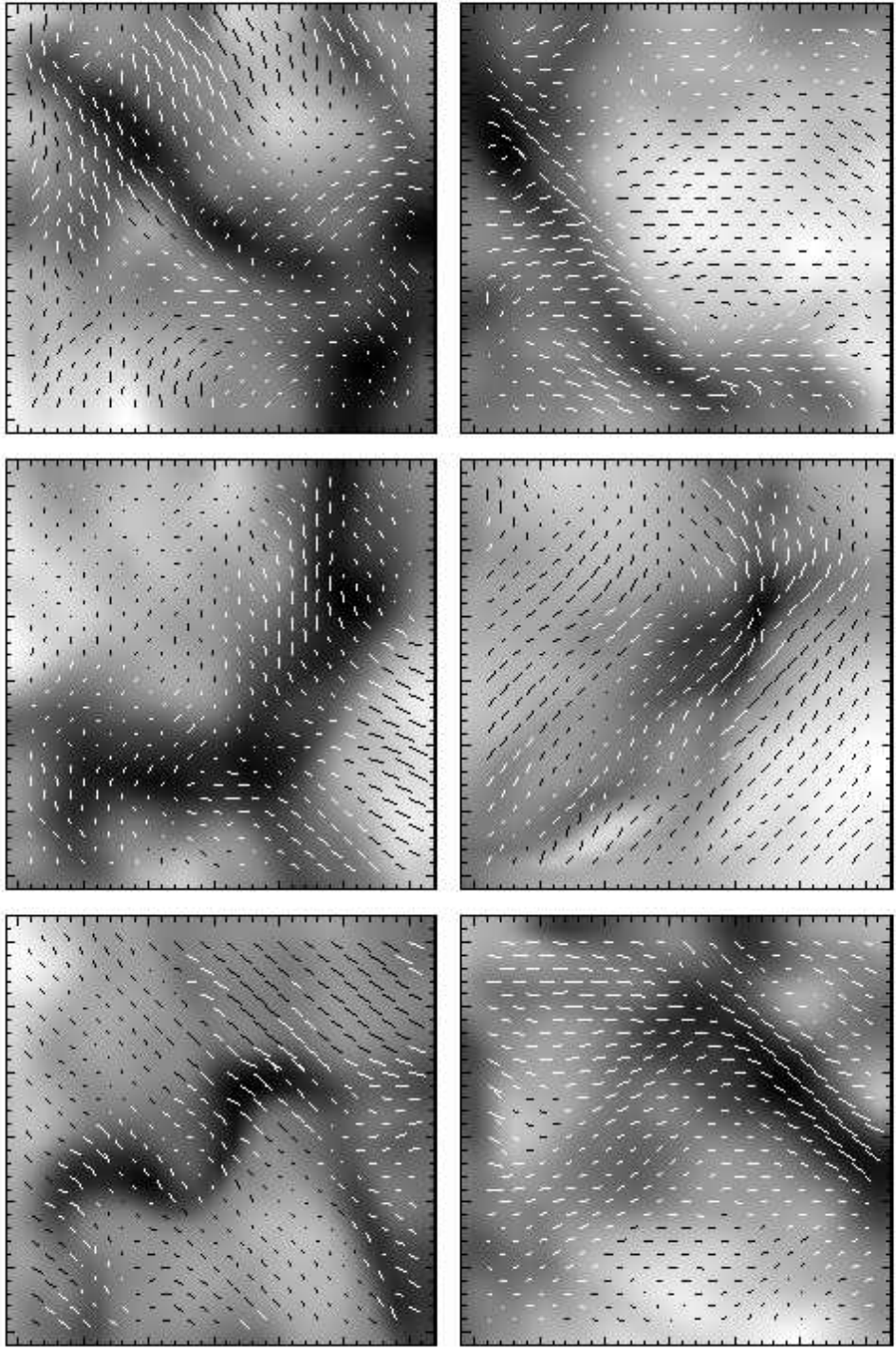}
\end{center}
\caption{\label{fig:highresfil}
%fig-filweak.ps
Magnetic polarization vectors and surface density for six selected filaments of model
${\cal E}r1a$ as denoted in Figure \ref{fig:highresmap}.}
\end{figure*}

In the simulations presented here, the filaments are solely due to shock
interactions. We do not find a preferred alignment of the magnetic field with
the filaments. We illustrate this in Figure \ref{fig:highresmap}, which shows 
the full polarization map for the $512^3$ zone model ${\cal E}r1a$, and with
a selection of filaments of the same model in Figure \ref{fig:highresfil}.
A detailed study of the filaments will be presented in a forthcoming paper.

%%%%%%%%%%%%%%%%%%%%%%%%%%%%%%%%%%%%%%%%%%%%%
%%%%%%%%%%%%%%%%%%%%%%%%%%%%%%%%%%%%%%%%%%%%%
% subsection{Estimating the Field Strength:...}
%%%%%%%%%%%%%%%%%%%%%%%%%%%%%%%%%%%%%%%%%%%%%
%%%%%%%%%%%%%%%%%%%%%%%%%%%%%%%%%%%%%%%%%%%%%
\subsection{Estimating the Field Strength: Chan\-drasekhar-Fermi-Method
            \label{subsec:res-cf}}

Chandrasekhar \& Fermi (1953; CF) suggested a method of estimating the mean 
magnetic field strength $\meanB$ in the galactic spiral arms.
It relates the line of sight velocity dispersion $\sigvlos$ to the dispersion of
polarization angles $\sigtand$ around a mean field component. The CF
method rests on two main assumptions: that the fluctuations are isotropic
about the direction of the mean field, and that there is equipartition
between the mean kinetic and magnetic energies in the fluctuations. Under
these conditions, the mean field component $\meanB$ is given by 
\begin{equation}
  \meanB^2 = 4\pi\rho \frac{\sigvlos^2}{\sigtand^2}
  \label{equ:cf-original}
\end{equation}
where $\rho$ stands for the mean density. Because of the angular variations
in the denominator, the result depends strongly on the actual mean field 
strength, and will in fact only be meaningful if there is a noticable mean
field component. In terms of the parameter 
$\xi$, the ratio of turbulent magnetic to turbulent kinetic energy listed in 
Table \ref{tab:turbcharacter}, equation (\ref{equ:cf-original}) generalizes to
\begin{equation}
  \meanB^2 = 4\pi\rho \frac{\sigvlos^2}{\sigtand^2}\xi.
  \label{equ:cf-equip}
\end{equation}
We initially set $\xi = 1$ when implementing the CF-method, because observations do not
yield clear evidence for or against equipartition (e.g. Crutcher 1999), but we
correct for $\xi \neq 1$ later (see Table \ref{tab:turbcharacter}).

In the following paragraphs, we present a modification of the CF-method
(\S~\ref{subsubsec:res-cf-method}), compare both methods with the help of
model data (\S~\ref{subsubsec:res-cf-calib}) and discuss the effect of limited telescope
resolution on the resulting field strength estimates in 
\S~\ref{subsubsec:res-cf-powerloss}.

%%%%%%%%%%%%%%%%%%%%%%%%%%%%%%%%%%%%%%%%%%%%%
% subsubsection{A Modification}
%%%%%%%%%%%%%%%%%%%%%%%%%%%%%%%%%%%%%%%%%%%%%
\subsubsection{A Modification\label{subsubsec:res-cf-method}}

The CF-method in its original form estimates the mean magnetic field. When
the mean field is much less than the rms field, the dispersion in fluctuation
angle $\sigma^2(\tan\delta)$ is dominated by points where $\delta\approx\pi/2$.
Since $\meanB\propto\sigma(\tan\delta)^{-1}$ (see eq.~(\ref{equ:cf-original})),
the result is usually an underestimate of $\meanB$.

The CF-method can be extended to yield an estimate
$\rmsB$ which is free of this problem. Suppose $\meanB = \hat z\meanB$ and
the $x-z$ plane is the plane of the sky. Then
\begin{equation}
  \rmsB = \meanB^2 + \langle \Delta B_{z}^2\rangle
                   + \langle B_x^2\rangle + \langle B_y^2\rangle.
  \label{equ:rms-b}
\end{equation}
By definition of the angle $\delta$
\begin{equation}
  B_x=\meanB \tan{\delta}.
  \label{equ:per-b}
\end{equation}

Squaring and averaging equation (\ref{equ:per-b}), assuming that the magnetic
field fluctuations energies are the same for all components 
(see \S~\ref{subsubsec:res-cf-claimtest}), and using equation
(\ref{equ:cf-original}), equation (\ref{equ:rms-b}) becomes
\begin{equation}
  \rmsB = 4\pi\rho \frac{\sigvlos^2}{\sigtand^2} (1+3\sigtand^2)
  \label{equ:cf-modified}.
\end{equation}

When the field is highly disordered, and the dispersion in polarization
angles is large, equation (\ref{equ:cf-modified}) reflects the underlying
physical assumptions of equipartition between kinetic and magnetic energy 
together with isotropy. The formula is valid for arbitrary ratios of turbulent
magnetic to turbulent kinetic energy if we multiply the RHS by $\xi$.
When the field is highly ordered, comparison of equation (\ref{equ:cf-modified}) 
with equation (\ref{equ:cf-original}) shows
that the mean and rms fields are nearly the same.

%%%%%%%%%%%%%%%%%%%%%%%%%%%%%%%%%%%%%%%%%%%%%
% subsubsection{Calibration with Model Data}
%%%%%%%%%%%%%%%%%%%%%%%%%%%%%%%%%%%%%%%%%%%%%
\subsubsection{Calibration with Model Data\label{subsubsec:res-cf-calib}}

\begin{figure*}
\begin{center}
\includegraphics[height=18.0cm]{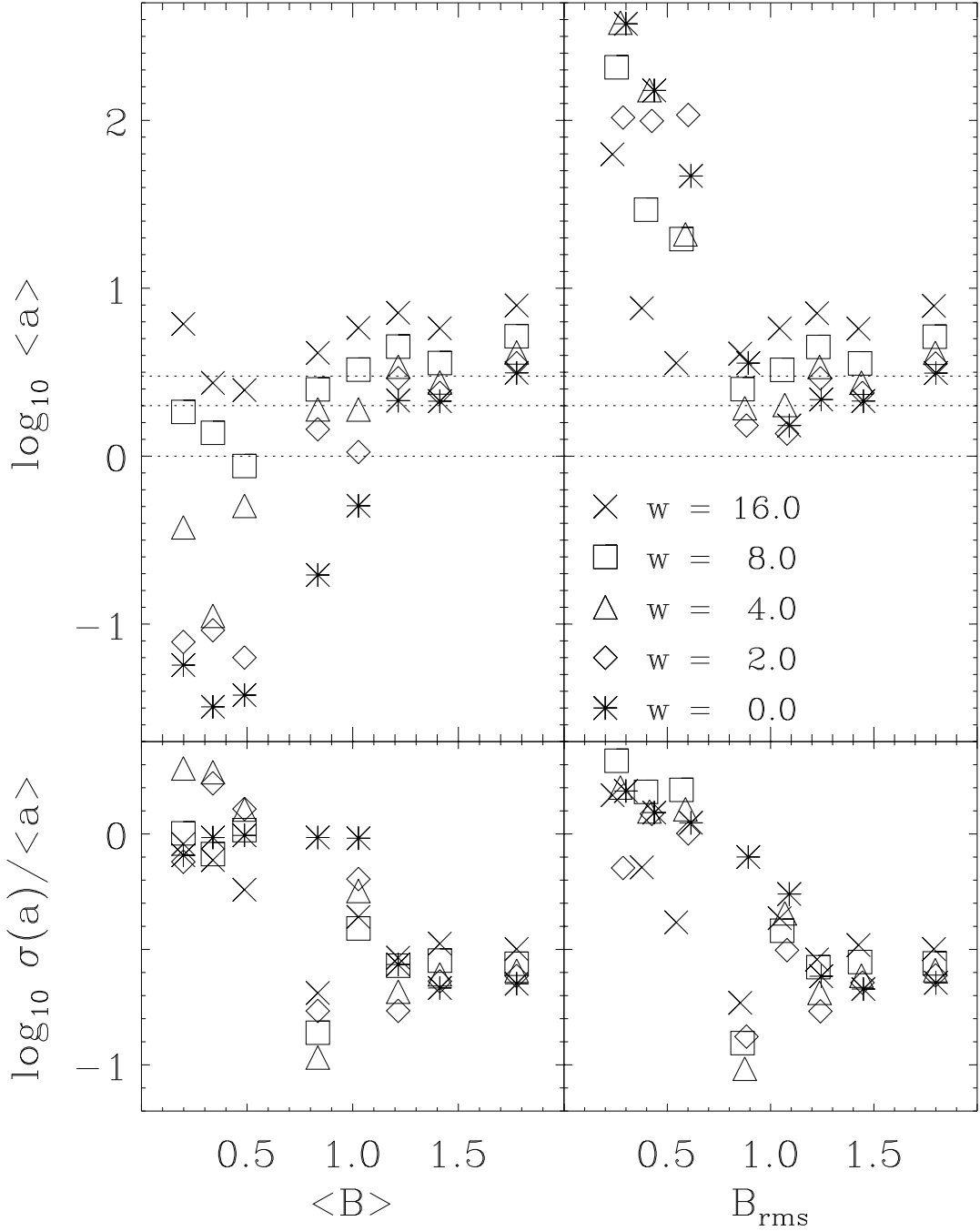}
\end{center}
\caption{\label{fig:cf-variation}
Overestimation factors $a_{CF}=B_{CF}/\meanB$ and $a_{rms}=B_{CF}^{mod}/B_{rms}$
for the CF-method in its original and modified version for all models
of type ${\cal G}$, averaged over two lines of sight and three physical times
(upper panels). The dotted lines correspond to $a=1,2,3$.
Symbols stand for smoothing beams applied. Increasing beam sizes lead to 
systematic overestimation of the field strengths. The beam size is given in
zones, with a total grid size of $128$ zones.
The lower panels contain the relative standard deviations with respect to the
corresponding mean value.}
\end{figure*}

We plotted the ratio of the estimated and model field strength
$a \equiv B_{est}/B_{model}$ in Figure \ref{fig:cf-variation}.
The upper left panel displays $a$ for the CF-method, averaged over two lines
of sight perpendicular to the mean field direction and over three
physical times. The lower left panel shows the dispersion of a single measurement
relative to the corresponding mean value. It decreases with increasing
field strength. For strong fields, the CF-method
overestimates the field strength derived from the unsmoothed maps
(star symbols in Fig. \ref{fig:cf-variation}) by a factor of $2$ to $3$. 
Thus we confirm the claim made by Ostriker et al. (2001).
For weaker fields however, the method starts to develop a significant 
scatter, as shown in Figure \ref{fig:cf-variation} (lower panel). 

The right column in Figure \ref{fig:cf-variation}
show the corresponding results for the CF-method in its modified version
according to equation (\ref{equ:cf-modified}). For the unsmoothed case
(again star symbols), the strongest field is overestimated by $a\approx 3$. 
Whereas the original method breaks down at a field strength of 
$B_{model}=1.2$, the modified version still yields acceptable results 
for $B_{model}=0.9$. 

The minimum field strength alone however does not tell us
much about the reliability of the method. The parameter of interest is the
Alfv\'{e}n Mach number ${\cal M}_A=\langle v^2\rangle^{1/2}/c_A$, as the angular variations 
$\sigtand$ of polarization not only depend on the energy content in the field, 
but on the turbulent kinetic energy as well. A field strength of $B_{model}=0.9$
would correspond to ${\cal M}_A = 1.25$, in the parameter set of models ${\cal G}$. 
With typical parameters of $c_s = 200\mbox{m}/\mbox{s}$ and $n(\mbox{H}_2) = 10^3\mbox{ cm}^{-3}$, 
we can then scale $B_{model} = 7\mu\mbox{G}$. Field strengths in molecular clouds 
are seen to be mostly larger than $B \gtrsim 10\,\mu\mbox{G}$ (Crutcher 1999). In denser 
regions, the field strengths seem to increase with density as $B \propto \rho^{0.47}$ 
(Crutcher 1999), whereas the nonthermal line widths (the ``turbulence'') decreases
with decreasing size as $\Delta v_{NT} \propto R^{0.21}$ (Caselli \& Myers 1995).
Thus we conclude, that both methods should yield acceptably accurate results for 
molecular cloud regions up to their densest parts, {\em as long as the observations
sample the angular variations sufficiently} (see \S~\ref{subsubsec:res-cf-powerloss}).

\begin{figure*}
\begin{center}
\includegraphics[height=18.0cm]{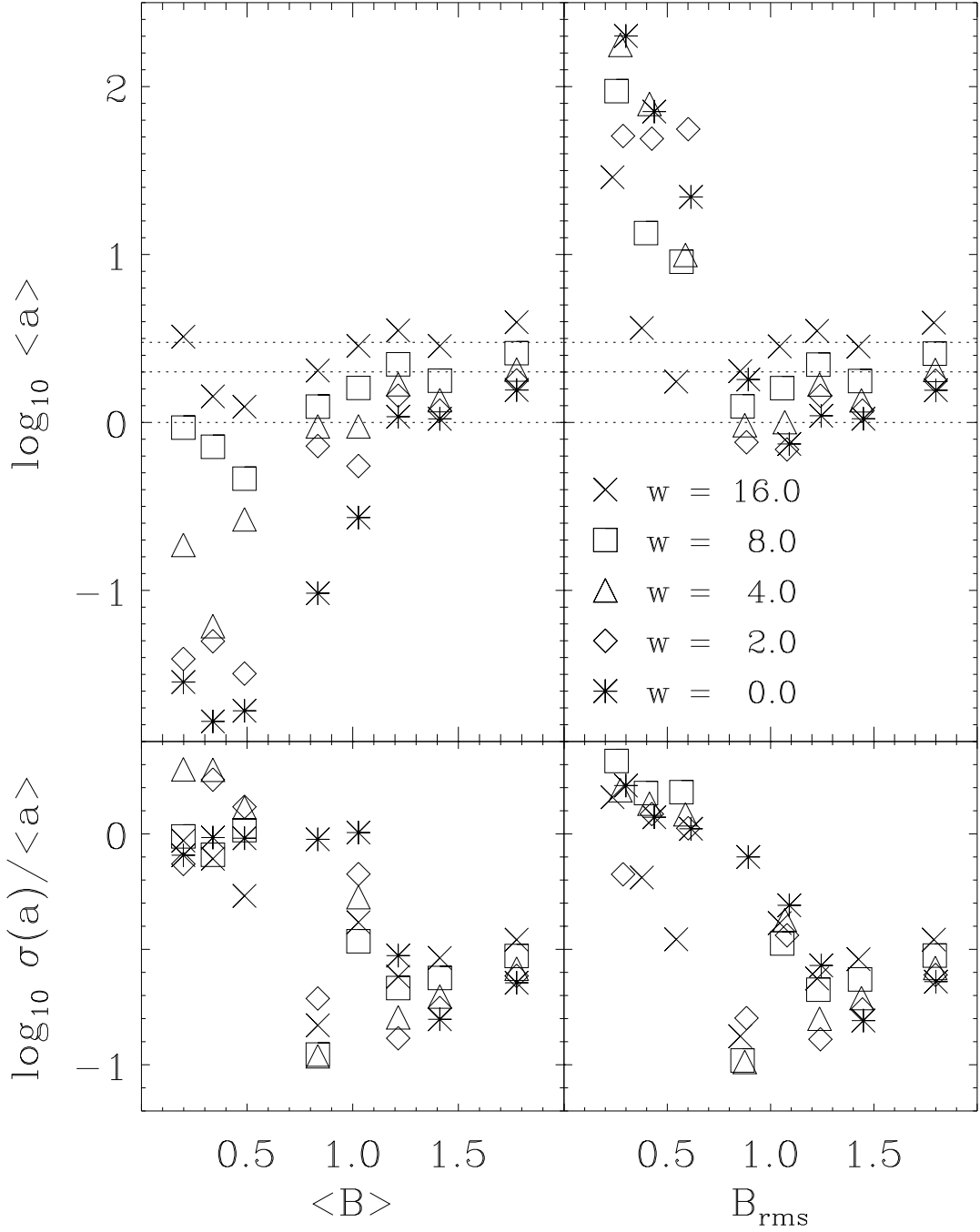}
\end{center}
\caption{\label{fig:cf-corrected}
Overestimation factors $a_{CF}=B_{CF}/\meanB$ and $a_{rms}=B_{CF}^{mod}/B_{rms}$
for the CF-method in its original and modified version for all models
of type ${\cal G}$ as in Figure \ref{fig:cf-variation}, but after correcting 
for non-equipartition (see Table \ref{tab:turbcharacter}) with $1/\sqrt{\xi}$. 
The dotted lines correspond to $a=1,2,3$. Both methods return
the model field strength nearly exactly for sufficiently high field strength
and no smoothing.}
\end{figure*}

We have to address the question of energy equipartition. Since $\xi$ is
generally less than unity in our models (see Table \ref{tab:turbcharacter}), 
the CF-method and its extension overestimate both $\meanB$ and $\rmsB$ by a 
factor of $1/\sqrt\xi$. Figure \ref{fig:cf-corrected} shows 
$a\equiv B_{est}/B_{model}$ ``corrected'' for non-equipartition. Both methods
now hit the model field strength at a factor between $1$ and $1.5$, at least for 
sufficiently large field 
strengths and for unsmoothed maps. The correction slightly reduces the largest 
deviations of the weak-field estimates. We conclude, that for a sufficiently well 
resolved (see below) polarization map and for large field strength, both methods 
yield reliable results. A main uncertainty factor is the ratio of magnetic to 
kinetic energy $\xi$ in the observed region.

Whereas the original CF-method underestimates low field strengths, the
modified version overestimates them. For weak fields, the polarization
angles can reach $90^o$ with respect to the mean field, which yields 
$|\tan \delta| = \infty$ in equations~(\ref{equ:cf-original}) and 
(\ref{equ:cf-modified}). 

As the original CF-method, the modified version yields overestimated
field strengths with increasing smoothing beam width.
From the lower row in Figure \ref{fig:cf-variation} we conclude that even for 
the strongest fields the relative scatter is of order $20$\%. 
Smoothing the maps leads to smaller scatter except for the weakest field strengths. 
  
\begin{figure*}
\begin{center}
\includegraphics[width=12.0cm]{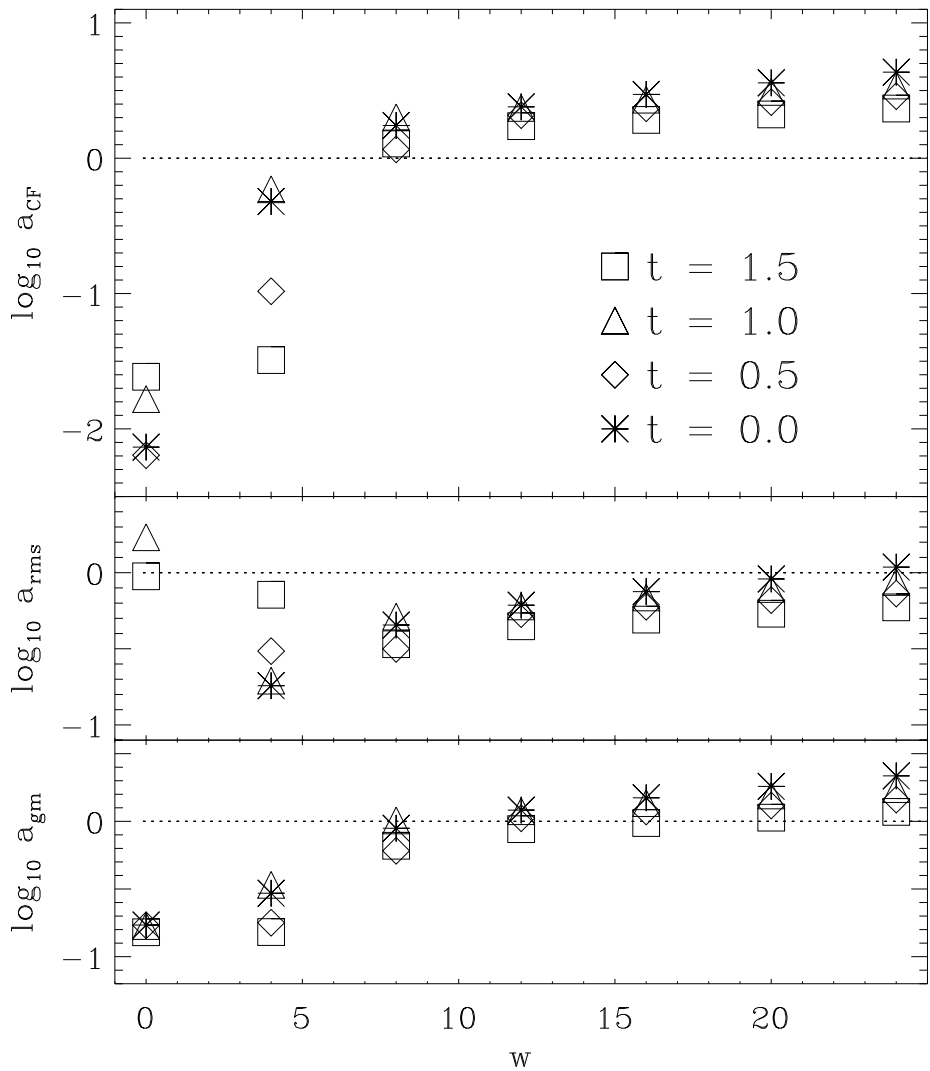}
\end{center}
\caption{\label{fig:cf-gravsigma}
Overestimation factors $a_{CF}=B_{CF}/\meanB$, $a_{rms}=B_{CF}^{mod}/B_{rms}$
and $a_{gm}=(a_{CF}a_{rms})^{1/2}$
for the self-gravitating model ${\cal E}h1d$. At $t=0.0$, gravity is switched on.
Note that all three panels are plotted in the same logarithmic scale.}
\end{figure*}

The effect of self-gravity on the field strength estimates is minute (Fig.
\ref{fig:cf-gravsigma}). For small smoothing widths $w$, the varying small scale
structure due to self-gravity leads to some statistical scatter, which is averaged out
for large $w$. The CF-methods seem to be insensitive to effects of self-gravity on
small scales.

%%%%%%%%%%%%%%%%%%%%%%%%%%%%%%%%%%%%%%%%%%%%%
% subsubsection{A Recipe}
%%%%%%%%%%%%%%%%%%%%%%%%%%%%%%%%%%%%%%%%%%%%%
\subsubsection{A Recipe
               \label{subsubsec:res-cf-recipe}}

\begin{figure*}
\begin{center}
\includegraphics[height=18.0cm]{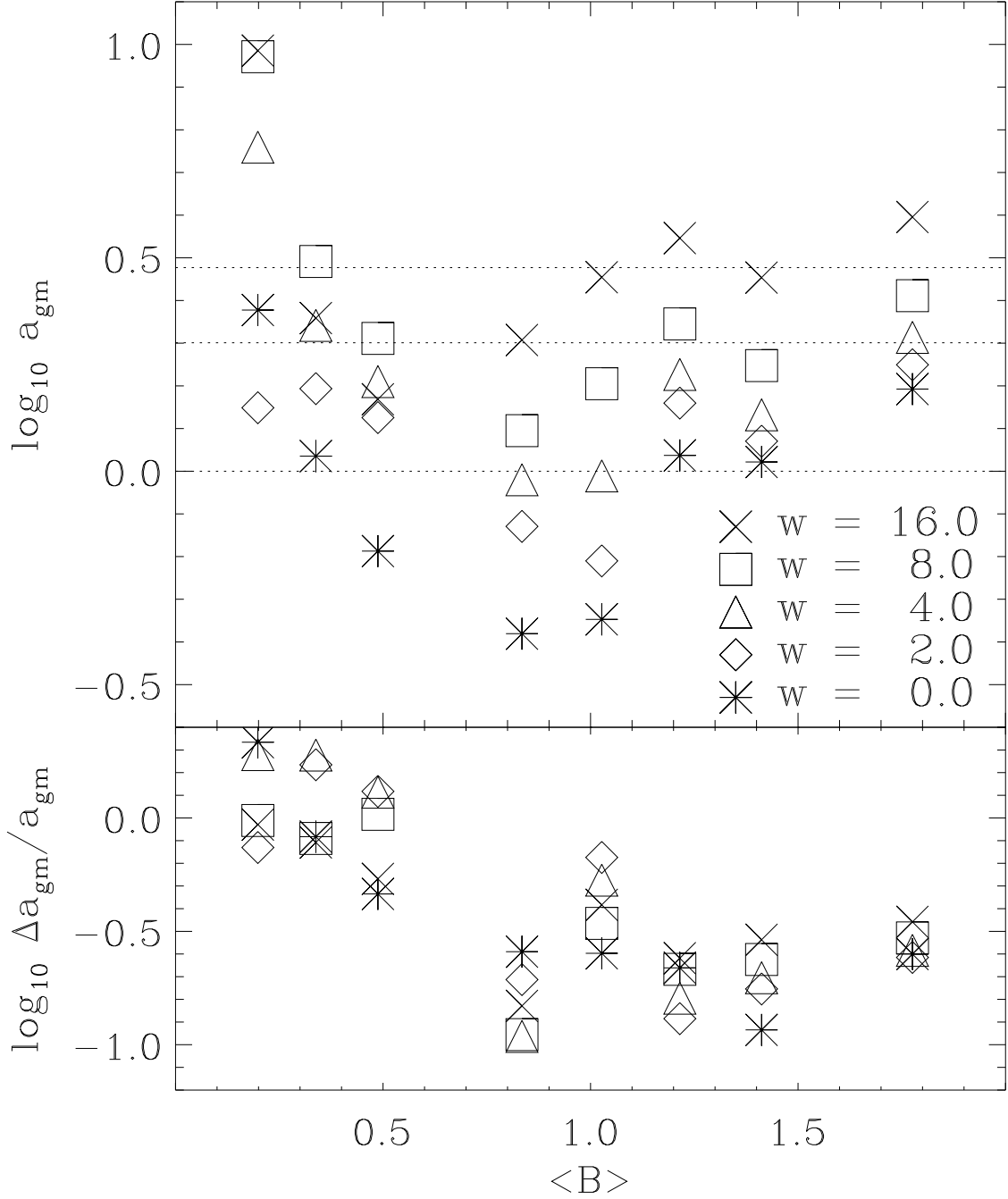}
\end{center}
\caption{\label{fig:cf-varcormean}
Geometric mean $a_{gm}=\sqrt{a_{CF}\,a_{rms}}$ against model field strength,
averaged and corrected for non-equipartition as in Figure~\ref{fig:cf-corrected}.
(upper panel). Where the separate estimates $a_{CF}$ and $a_{rms}$ 
(Fig.~\ref{fig:cf-corrected}) miss the weak field strengths by a factor
$ > 150$ for unsmoothed ($w=0$) maps, $a_{gm}$ yields an estimate accurate 
up to a factor of $\approx 2.5$ for the same $w=0$-maps. Smoothed fields
are still systemetically overestimated. 
The lower panel shows the corresponding errors.}
\end{figure*}

 The fact that $a_{CF}$ and $a_{rms}$ vary by roughly the same
logarithmic range, but with opposite signs, suggests that by taking the geometric
mean
\begin{equation}
  \begin{split}
    a_{gm} &= \sqrt{a_{CF}\,a_{rms}}\\
           &= 4\pi\rho \frac{\sigvlos^2}{\sigtand^2} (1+3\sigtand^2)^{1/2}
    \label{equ:res-cf-recipe-agm}
  \end{split}
\end{equation}
we would arrive at a more accurate estimate for the actual field strength.
Applying this recipe yields Figure~\ref{fig:cf-varcormean} (corresponding to
Fig.~\ref{fig:cf-corrected}).
Clearly, the large deviations of $a_{CF}$ and
$a_{rms}$ cancel sufficiently to yield estimates accurate to a factor of
$a_{gm} \approx 2.5$ even for the weakest, unsmoothed fields, and is more accurate than that
for moderately strong fields. Smoothed maps of course lead again to a systematic 
overestimation, here up to a factor of $10$. We would like to emphasize that
equation~(\ref{equ:res-cf-recipe-agm}) is only motivated by the (logarithmically) 
comparable deviations of $a_{CF}$ and $a_{rms}$.

\begin{figure*}
\begin{center}
\includegraphics[height=18.0cm]{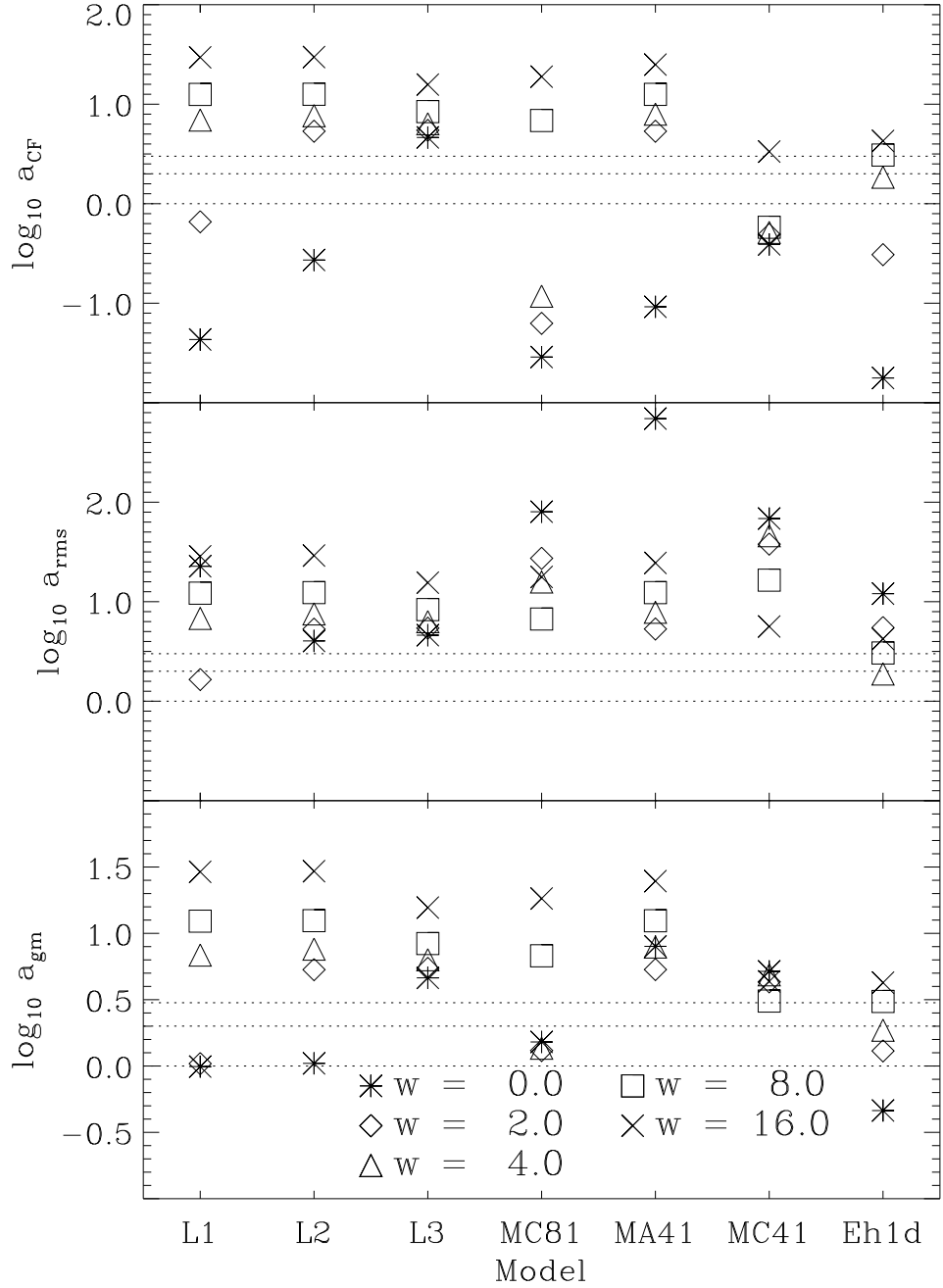}
\end{center}
\caption{\label{fig:cf-vcrest}
$a_{CF}$, $a_{rms}$ and geometric mean $a_{gm}=\sqrt{a_{CF}\,a_{rms}}$ for an
extended model set (see Table~\ref{tab:models}, lower part),
corrected for non-equipartition as in Figure~\ref{fig:cf-corrected}. 
Where the separate estimates $a_{CF}$ and $a_{rms}$ scatter around $a=1$ by
up to a factor of $700$ for unsmoothed ($w=0$) values, $a_{gm}$ reduces 
this scatter to a factor $< 10$.}
\end{figure*}

To test this recipe, we applied it to a set of models with varying physical
properties (Fig.~\ref{fig:cf-vcrest}). Models $L1$, $L2$ and $L3$ are a
time series of decaying turbulence taken from Mac Low (1999). Models
$MC81$, $MA81$ and $MC41$ (Mac Low et al. 1998) are simulations of driven turbulence as
series ${\cal G}$, but with driving wave length of $k=3-4$ ($MC41$) and 
$k=7-8$ ($MA81$, $MC81$). See Tables~\ref{tab:models} and \ref{tab:turbcharacter}
for their parameters. All these models have weak initial fields, corresponding
to ${\cal G}i1b$. As in Figure~\ref{fig:cf-varcormean}, the scatter of $a_{gm}$
is more than one order of magnitude lower than for $a_{CF}$ and $a_{rms}$.
The unsmoothed estimates are accurate again up to a factor of $a_{gm} \approx 2.5$.
Again, smoothed fields are likely to be overestimated. Note that $a_{rms}$ on the
whole leads already to slightly more accurate estimates than $a_{CF}$. 

We conclude that the geometric mean $a_{gm}$ gives more reliable estimates
than $a_{CF}$ or $a_{gm}$ even for very weak fields, being more accurate as well
for moderate and strong fields. Thus, this recipe may prove useful when being
applied to observational data.

%%%%%%%%%%%%%%%%%%%%%%%%%%%%%%%%%%%%%%%%%%%%%
% subsubsection{Smoothing, subsampling and...}
%%%%%%%%%%%%%%%%%%%%%%%%%%%%%%%%%%%%%%%%%%%%%
\subsubsection{Smoothing, subsampling and their effect in a turbulent medium
               \label{subsubsec:res-cf-powerloss}}

Field estimates derived from smoothed maps tend to overestimate the
field strength with increasing ``beam width'' (Fig. \ref{fig:cf-variation}, 
all symbols except stars). We will discuss how limited telescope resolution
in polarization maps of turbulent regions can lead to this overestimation.

Both CF-methods rely on perturbations of the magnetic field around 
a mean component. If all the power of these perturbations resided
on spatial scales $1/\hat{k}=\hat{w}<w$, increasing the smoothing  
width above $w$ would have no effect on the power distribution of 
polarization angles, and varying smoothing widths would yield identical results. 
On the other hand, if the perturbations grew with increasing length scale, 
as e.g. in a turbulent power spectrum, a larger smoothing width $w$ would 
result in a greater power loss on larger scales, and thus to an overestimation 
of the field. Note that in equations~(\ref{equ:cf-original}) and (\ref{equ:cf-modified}) 
we find the angular variations in the denominator. So losing power in
the variations increases the field estimate. The turbulence is driven 
between wave numbers $k=1-2$ in our simulations, thus letting the code evolve 
a turbulent cascade down to the dissipation scale at grid cell size. 
This turbulent cascade in our models is well reflected in the power spectrum 
of angular variations $\sigma(\delta)$ (Fig.~\ref{fig:smoothphi}).
We lose more and more power when applying larger smoothing widths $w$. 
This power loss should depend on the steepness of the power spectrum. 
This suggests that it might be possible to measure the spectrum of magnetic
field fluctuations by smoothing polarization maps by different filter widths,
and comparing the dispersion of polarization angles.

\begin{figure*}
\begin{center}
\includegraphics[width=12.0cm]{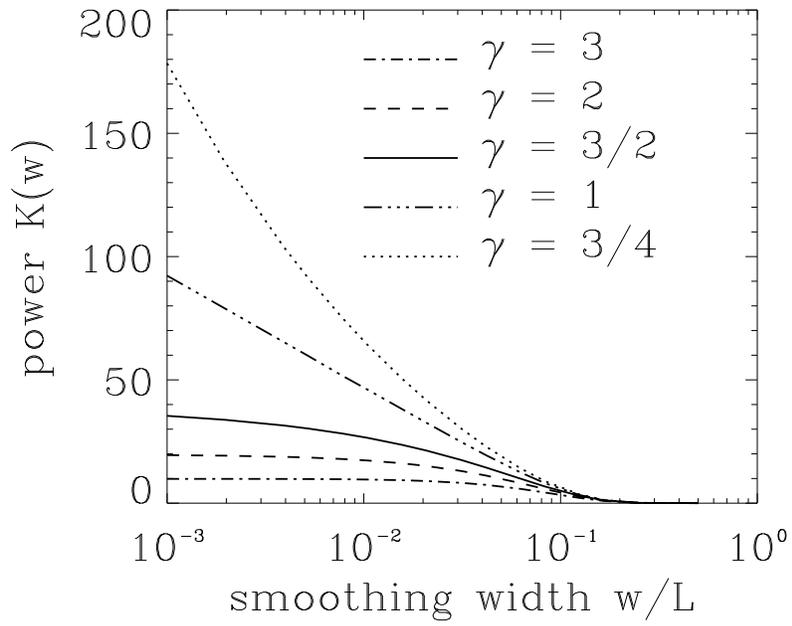}
\end{center}
\caption{\label{fig:cf-powerloss}
Analytic results for the amount of power remaining in the spectrum $K(w)$ (see
eq.~(\ref{equ:spectrum-keep})) for five values of the power law exponent
$\gamma$. In this log-linear plot, exponents of $\gamma > 1$ and $\gamma < 1$ can
be distinguished by their curvature. $L$ gives the box length.}
\end{figure*}

\begin{figure*}
\begin{center}
\includegraphics[width=12.0cm]{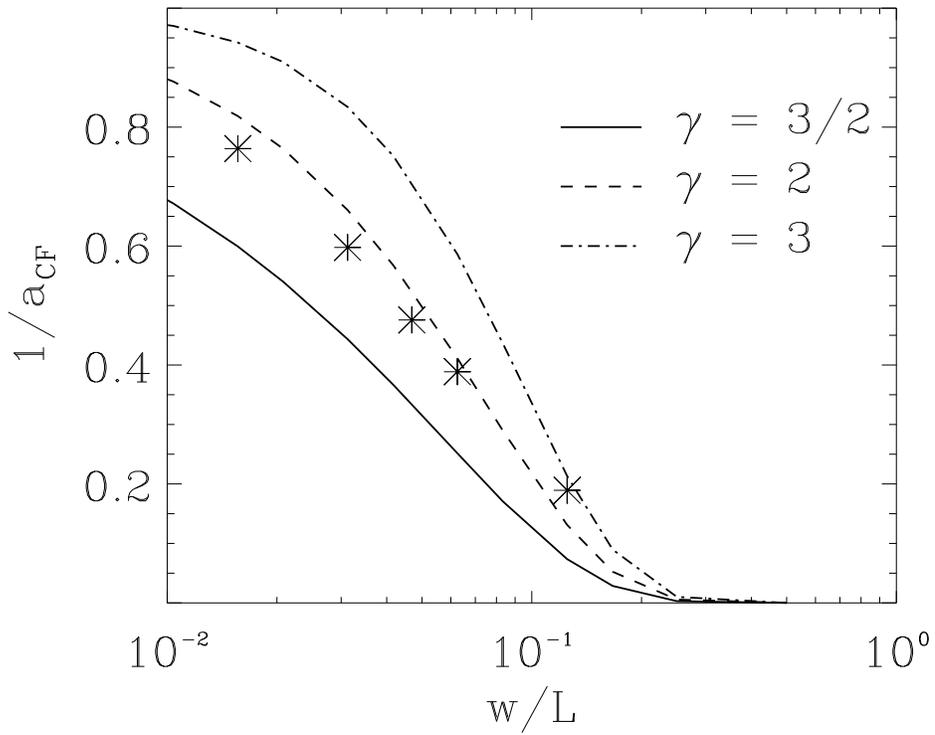}
\end{center}
\caption{\label{fig:cf-powerlosscomp}
Analytic results for the amount of power remaining in the spectrum $K(w)$ (see
eq.~(\ref{equ:spectrum-keep})) for three values of the power law exponent
$\gamma$. The stars denote the inverse of $a_{CF}$ 
taken from model ${\cal G}h1h$. All curves have been normalized to
$w=0$ in order to compare them. $L$ is the box length.}
\end{figure*}

To illustrate this relation, we will describe the smoothing as a convolution
and then will determine analytically the loss of power in a spectrum due
to smoothing. The convolution would be
\begin{equation}
  {\cal A}(x') = \int_{-\infty}^{\infty} A(x) F(x-x')\,dx
  \label{equ:def-convolution}
\end{equation}
with the wave function $A(x)$ 
\begin{equation}
  A(x) = \sum_{k=1}^{\infty} \left(\frac{1}{k}\right)^\gamma \sin(2\pi kx)
  \label{equ:def-wave}
\end{equation}
and the Gaussian filter function $F(x-x')$ as in equation~(\ref{equ:gaussfilter}).
We identify $A(x)$ with the $B_x^2/\langle B\rangle ^2$ of equation~(\ref{equ:per-b}).
Instead of convolving directly,
we use the Fourier transforms ${\cal F}[A]$ and ${\cal F}[F]$ of
equations~(\ref{equ:def-wave}) and (\ref{equ:gaussfilter}) to derive the power spectrum
of the convolved function $|{\cal F}[{\cal A}]|$ via the convolution theorem
\begin{equation}
  \hat{\cal A} = \hat{A} \hat{F},
  \label{equ:theorem-convolution}
\end{equation}
where the hatted quantities denote Fourier transformations.
The Fourier transform of a Gaussian yields again a Gaussian, in our case with
an additional phase shift. The frequency space variable is $\omega$.
\begin{equation}
  \begin{split}
    \hat{F}(\omega) &= \frac{1}{\sqrt{2\pi}w}
                       \int_{-\infty}^{\infty}
                       \exp\left(-\frac{1}{2}\left(\frac{x-x'}{w}\right)^2\right)
                       e^{-i\omega x}\,dx\\
                    &= e^{i\omega x}\,\exp\left(-\frac{1}{2}(\omega w)^2\right)
  \end{split}
  \label{equ:fou-filter}
\end{equation}
Applying the Fourier transform to the wave function $A(x)$ and integrating
over $k=[1,\infty]$, we get
\begin{equation}
  \begin{split}
    \hat{A}(\omega) &= \int_{-\infty}^{\infty}
                       \int_{k=1}^{\infty}\left(\frac{1}{k}\right)^\gamma
                       \sin(2\pi kx)
                       e^{-i\omega x}\,dk\,dx\\
                    &=i\pi \left(\frac{2\pi}{\omega}\right)^\gamma
  \end{split}
  \label{equ:fou-wave}
\end{equation}
Now we can multiply $\hat{A}$ and $\hat{F}$, and taking the absolute value gives
us the power spectrum of the convolved function $\cal{A}$.
\begin{equation}
  |\hat{\cal{A}}| = \pi \exp\left(-\frac{1}{2}(\omega w)^2\right)
                           \left(\frac{2\pi}{\omega}\right)^\gamma
  \label{equ:powerspectrum}
\end{equation}
As we are interested in the loss of power caused by the smoothing, we now
try to determine the amount of power remaining in the spectrum after smoothing,
i.e. the quantity
\begin{equation}
  K(w) = \int_{0}^{\infty} \pi\,\exp\left(-\frac{1}{2}(\omega w)^2\right)
         \left(\frac{2\pi}{\omega}\right)^\gamma \,d\omega
  \label{equ:spectrum-keep}
\end{equation}
For $\gamma \leq 1$, $\lim_{w\rightarrow 0}K(w) = \infty$. Figure \ref{fig:cf-powerloss}
illustrates the behaviour of $K(w)$ for several values of $\gamma$. Note that this 
is a log-linear plot, so that, for example, $K(w,\gamma=3/4) \propto \log(w)^2$. 
Curves for $\gamma > 1$ have negative, the ones for $\gamma < 1$ positive curvature.
Normalizing the curves for $\gamma < 1$ with $K(0)$ and plotting them together 
with the inverse of the ratio of estimated to model field strength $a_{CF}$
of model ${\cal G}h1h$ yields Figure \ref{fig:cf-powerlosscomp}. We normalized
$a_{CF}$ to $w=0$ as well, in order to compare the measurements to the analytic
curves $K(w)$. We note that the measurements start to flatten off for small $w/L$ 
because of line of sight averaging, which we do not include
in the analytic derivation. The same argument applies for observations. Thus, even 
in the ideal case of $w=0$, the measurement may not reach $a = 1$ but may overestimate 
the actual field strength. Of course, this would crucially depend on the length of 
line of sight and, generally, on optical depth effects. They, however, are neglegible
in the far-infrared regime. Because we need a certain length in the line of sight
in order to derive a meaningful velocity dispersion $\sigvlos$, this overestimate
due to line-of-sight averaging gives an intrinsic error in the CF-methods.
The power spectrum of angular variations
in the simulations is by no means a clear power law, due to numerical dissipation and the 
signatures of the driving scale. Thus, the measured $1/a_{CF}$ (Fig. 
\ref{fig:cf-powerlosscomp}) does not correspond clearly to one of the analytic curves.
However, if observations show a well defined power law in the angular variations $\sigtand$, 
and provide sufficient resolution to create  a series of maps smoothed over larger 
and larger areas, then it should be possible to extract 
information about the exponent of the power law and to correct the field
estimate as well. Once a power law description is found for the observed region, $K(w)/K(0)$ 
supplies a correction factor for the field strength estimate $B_{est}$. 

\begin{figure*}
\begin{center}
\includegraphics[width=12.0cm]{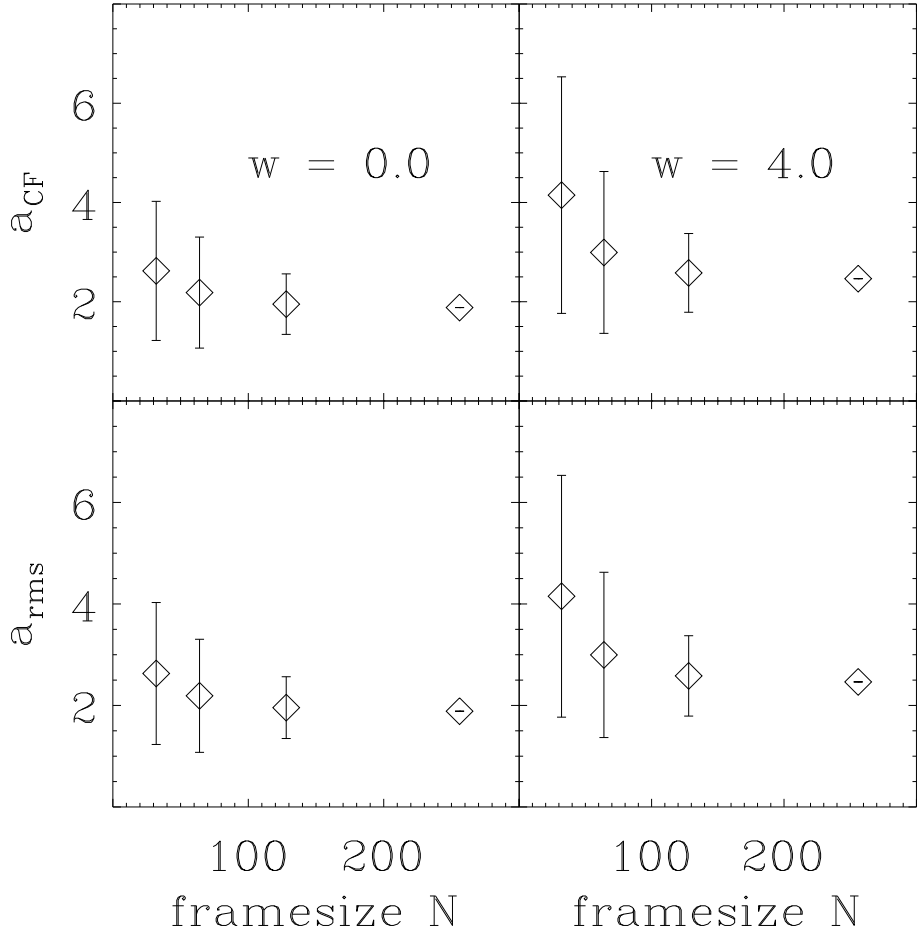}
\end{center}
\caption{\label{fig:cf-subframes}
%fig-cf-subframes.ps
Overestimation factors $a_{CF}=B_{CF}/\meanB$ and $a_{rms}=B_{CF}^{mod}/B_{rms}$
against linear size of subframes for model ${\cal G}h1h$. $a_{CF}$ and $a_{rms}$ were
determined by averaging over all possible subframes for a given box size.}
\end{figure*}

In polarization maps derived from observations, an additional effect could
potentially lead to a systematic overestimation.
When we derive the field estimates from our models, we use the complete domain, 
thus including all wave lengths occurring in the problem. However, it is not clear
whether the observations do trace the largest scales in the field perturbations.
If the power of angular variations is distributed in a power spectrum which 
decreases with increasing wave number, we will lose overproportionally much power 
when estimating the field strength within a domain which is smaller than the largest 
wave length contributing to the perturbations.
Figure \ref{fig:cf-subframes} illustrates this effect contributing
to a systematic overestimation of the field strength. It shows the ratios
$a=B_{est}/B_{model}$ for subframes of $32^2$, $64^2$ and $128^2$ cells for model
${\cal G}h1h$ at $256^3$ resolution. We averaged the estimates from all possible
subframes. The error bars indicate the standard deviations of the means. 
With decreasing frame size, the field is overestimated slightly, thus indicating
a decrease of $\sigtand$. However, bearing in mind the large scatter around the 
mean values in Figure \ref{fig:cf-subframes}, this effect is not very distinct.
   
We conclude that, although the CF-methods are applicable in principle to molecular
clouds, the resulting field strengths are most likely to be systematically 
overestimated for two reasons: (a) Any smoothing introduced by the finite telescope 
beam size will lead to a systematic removal of large-scale perturbations and thus to 
an overestimation of the field strength. (b) Deriving field estimates for a domain  
smaller than the maximum perturbation wave length present leads to power loss
on large scales and thus as well to an overestimated field strength.
This systematic overestimation due to limited resolution is different from uncertainties
because of non-equipartition as discussed in \S~\ref{subsubsec:res-cf-calib}.
In order to draw these conclusions we assume that the underlying power spectrum of 
perturbations follows a power-law-like distribution, at least, that it decreases with 
increasing wave number.

%%%%%%%%%%%%%%%%%%%%%%%%%%%%%%%%%%%%%%%%%%%%%
% subsubsection{Resolution Test}
%%%%%%%%%%%%%%%%%%%%%%%%%%%%%%%%%%%%%%%%%%%%%
\subsubsection{Resolution Test\label{subsubsec:res-cf-resolution}}

\begin{figure*}
\begin{center}
\includegraphics[width=12.0cm]{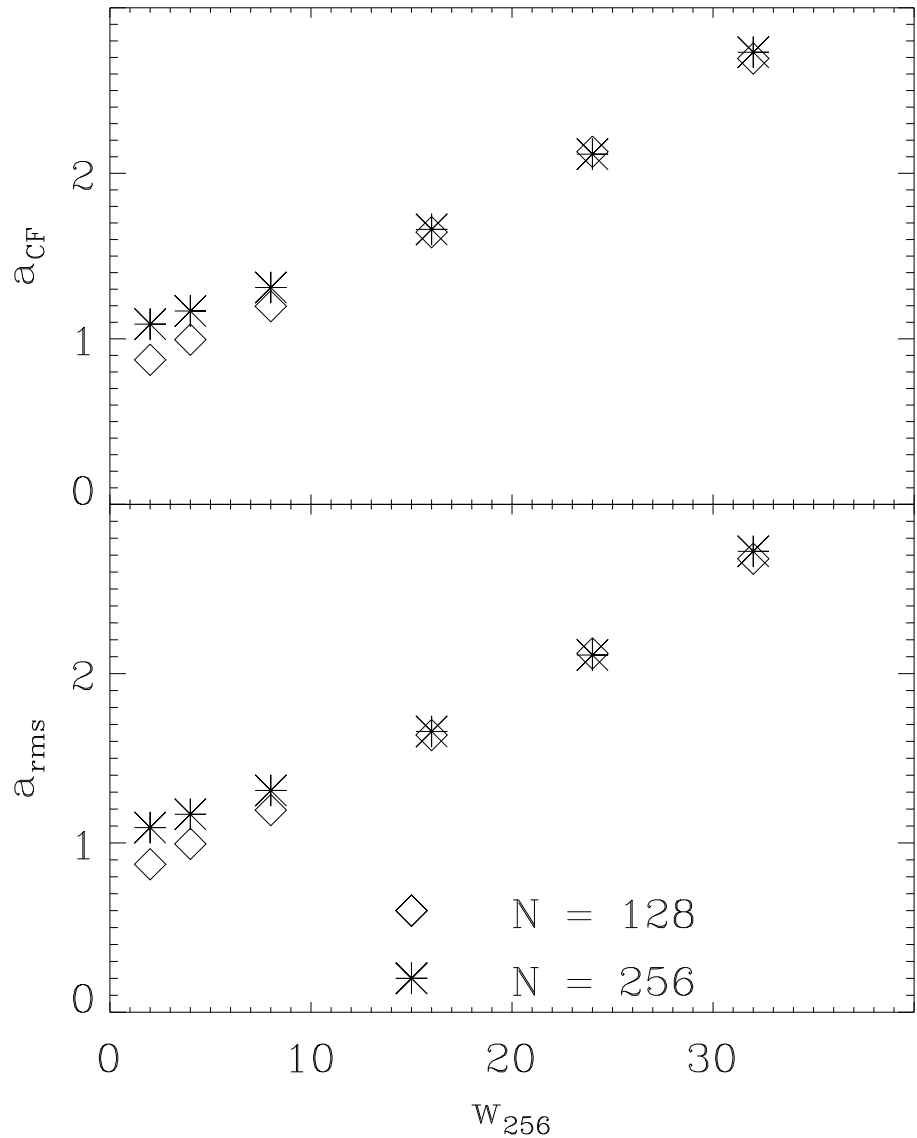}
\end{center}
\caption{\label{fig:resolution-cf}
%fig-resolution-cf.ps
Overestimation factors $a_{CF}=B_{CF}/\meanB$ and $a_{rms}=B_{CF}^{mod}/B_{rms}$
against beam width $w$ for the strong-field models ${\cal G}i1h$ and ${\cal G}h1h$ 
at resolutions of $128^3$ and $256^3$ cells. The beam width $w_{256}$ is given in 
pixels, where $w_{256}$ refers to the beam width of model ${\cal G}h1h$. 
We have to compare the different resolutions at the same physical beam width, which is
why the data points of model ${\cal G}i1h$ are shifted by a factor of $2$, i.e.
$w_{128} = 2 \rightarrow w_{256} = 4$.
}
\end{figure*}

Figure \ref{fig:resolution-cf} presents a test for resolution with 
models ${\cal G}i1h$ and ${\cal G}h1h$ at $128^3$ and $256^3$ grid cells.
The beam width $w_{256}$ is given in pixels. Note that in order to compare
both models, we have to compare them at the same {\em physical} beam width,
i.e. we have to shift a data point of model ${\cal G}i1h$ at $w_{128}=4$ 
such that it corresponds to a point of $w_{256}=8$.
We corrected all estimates with $1/\xi$ (see Table \ref{tab:turbcharacter}) 
in order to account for non-equipartition.
Discrepancies at small scales are to be expected as model ${\cal G}i1h$ then
already reaches the grid scale. Figure \ref{fig:resolution-cf} allows us to
conclude that numerical resolution does not affect our calibration of the
CF-methods. Thus we feel justified to draw conclusions for observations
of turbulent regions from our simulated observations of computational
turbulence.

%%%%%%%%%%%%%%%%%%%%%%%%%%%%%%%%%%%%%%%%%%%%%
% subsubsection{Isotropy of simulated turbulence}
%%%%%%%%%%%%%%%%%%%%%%%%%%%%%%%%%%%%%%%%%%%%%
\subsubsection{Isotropy of simulated turbulence\label{subsubsec:res-cf-claimtest}}

\begin{figure*}
\begin{center}
\includegraphics[width=12.0cm]{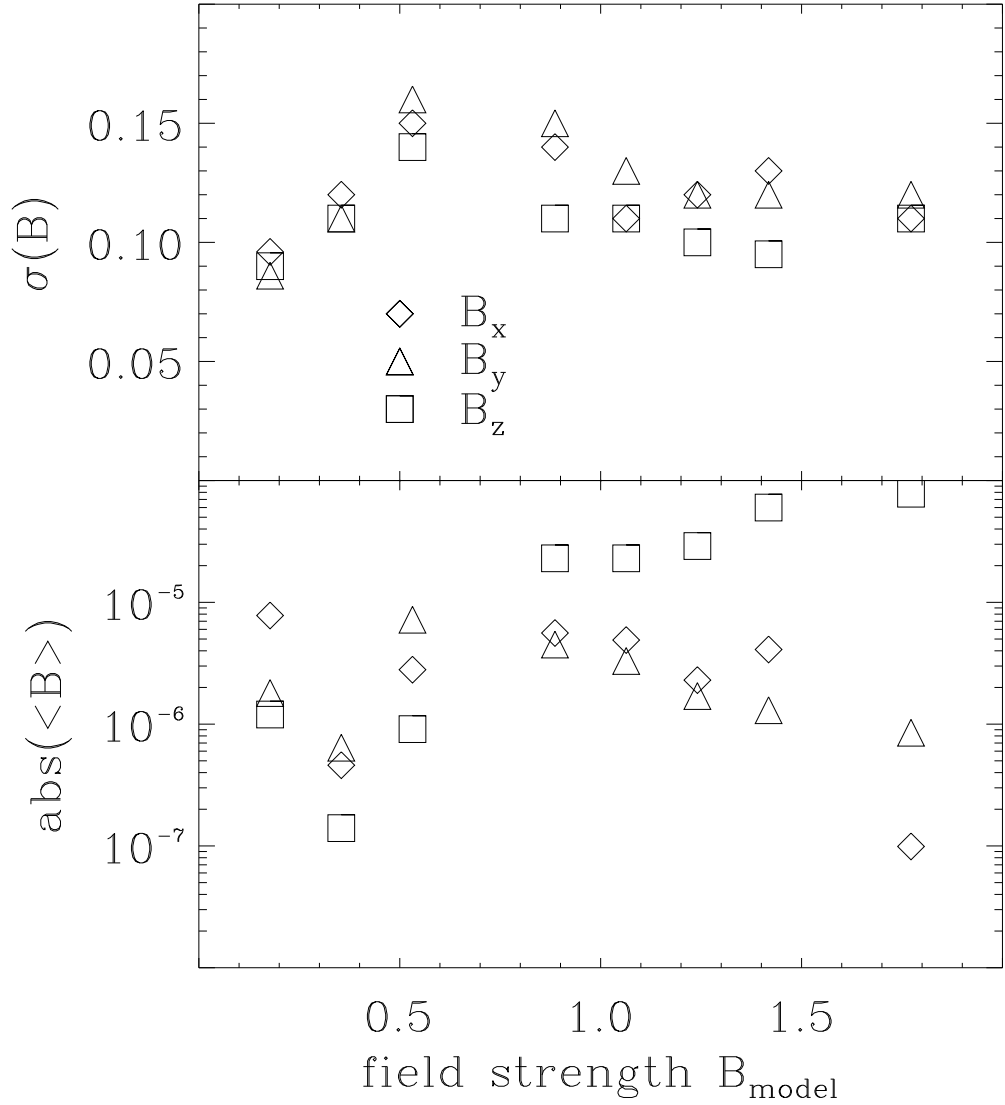}
\end{center}
\caption{\label{fig:cf-claim-test}
{\em rms} field values (upper panel) and mean values (lower panel) of
the magnetic fields in the three coordinate directions $x$, $y$ and $z$.
The initially uniform field is parallel to the $z$-direction. For the 
$z$-direction, we determined the mean value after subtraction
of the background field.}
\end{figure*}

We still have to substantiate the claim leading to equation~(\ref{equ:rms-b}),
namely, that our simulated turbulent flows are isotropic enough to validate
that the field components perpendicular to the mean field 
$\langle B\rangle = \hat{z} \langle B\rangle$ obey
$\langle B_x\rangle = \langle B_y\rangle$  and that the field perturbations in
all three directions fulfill
$\langle \Delta B_{z}^2\rangle = \langle B_x^2\rangle = \langle B_y^2\rangle$.
The upper panel of Figure \ref{fig:cf-claim-test} supports the second 
assumption. There occur only minor differences in the {\em rms} values.
Note that the initially homogeneous background field is oriented along the
$z$-direction. Thus, we had to subtract this background in order to determine
the corresponding mean values, yielding $\langle B_{z} \rangle$ (lower panel of 
Fig. \ref{fig:cf-claim-test}).
The scatter of the means is considerably larger, however their absolute values 
are three to four orders of magnitudes lower than the corresponding field strengths. 
Ideally, the means should vanish, especially in the directions perpendicular to
the initial field, namely the $x$- and $y$-direction. Deviations from this ideal
case are probably due to numerical diffusion in the code. This results in not
conserving the total flux per coordinate direction. The small shifts in the means
thus indicate how well the code actually does conserve the total flux.

%%%%%%%%%%%%%%%%%%%%%%%%%%%%%%%%%%%%%%%%%%%%%
%%%%%%%%%%%%%%%%%%%%%%%%%%%%%%%%%%%%%%%%%%%%%
%
% section{Conclusions and Implications}
%
%%%%%%%%%%%%%%%%%%%%%%%%%%%%%%%%%%%%%%%%%%%%%
%%%%%%%%%%%%%%%%%%%%%%%%%%%%%%%%%%%%%%%%%%%%%
\section{Conclusions and Implications\label{sec:conclusions}}

Reliable methods for measuring and mapping magnetic fields in molecular clouds
are urgently needed in order to assess the role of magnetic fields in the
dynamics of molecular clouds and in star formation. Probes based on far 
infrared polarimetry of the thermal emission from magnetically aligned 
dust are now in widespread use. Far-IR polarimetry 
is complementary to the Zeeman
method, which is the traditional means of measuring the field. There are few
independent checks of its accuracy as a diagnostic of magnetic fields.

In this paper we used numerical simulations of turbulent, magnetized, self
gravitating molecular clouds to assess the reliability of polarization maps.
Although such simulations qualitatively reproduce many of the features seen
in molecular clouds, we do not expect the computed densities, velocities,
and magnetic fields to be completely realistic. We undertook the project with
the expectation that analysis of simulated data would bring out the same
issues as analysis of real data. 

We assumed that the polarized emission is optically thin and
proportional to gas density. The first assumption should be excellent. The
second could be modified, for example by modelling the grain temperature
point to point within the cloud. It might be possible to pick out particular
structures along the line of sight by comparing maps at different wavelengths,
but establishing this is beyond the scope of the present paper.

The salient properties of the models are given in Tables \ref{tab:models} 
and \ref{tab:turbcharacter}. Based on these models, our principal conclusions 
are as follows:

\begin{enumerate}
  \item Filaments produced by shocks do not show a preferred alignment with the 
        magnetic field. 

  \item Self-gravity has no discernible effect on the structure of
        the magnetic fields in our simulations. However, we have to bear mind
        that our self-gravitating regions are small with respect to the
        simulated region.

  \item The Chandrasekhar-Fermi method in its original version yields field
        estimates accurate up to $a_{CF} = B_{CF}/\meanB = 2$.
        A modified version determining the rms field $\Bcfmod$ results in
        in $a_{rms} = \Bcfmod/\Brms = 2$. This version proves to work more reliably
        for slightly weaker field strengths. Both methods should be applicable to magnetic
        fieldstrengths typical of molecular clouds.

  \item The geometrical mean $a_{gm} = \sqrt{a_{CF}\,a_{rms}}$ yields field estimates
        accurate up to a factor of $\approx 2.5$ even for the weakest fields.
        For moderate and stronger fields, $a_{gm} \approx 1$. Thus, this recipe
        improves the reliability of field strength estimates for the whole range of 
        physically realistic fields in molecular clouds, and may prove useful when applied 
        to observational data. 
        
  \item The original and modified Chandrasekhar-Fermi methods are based on
        the assumption of equipartition between turbulent kinetic and magnetic
        energy. The field fluctuations in the models are below equipartition, 
        leading to an overestimate of the field.

  \item Limited angular resolution again leads to systematic overestimation of the magnetic
        field strength for both methods. Two effects can cause this overestimation:
        (a) In a power spectrum decreasing with increasing wave number, smoothing
        leads to an overproportionally large power loss on larger scales.
        Thus the angular variations are underestimated, which leads to larger 
        field strengths. (b) If the domain used for deriving the field estimate
        does not trace the largest wave modes, this again leads to power loss on
        the largest scales. 

  \item If the power spectrum of magnetic field fluctuations is a power
        law, field estimates made from a series of maps with different smoothing
        widths can in principle	
        be used to estimate the exponent of the fluctuation spectrum, assuming that
        this is well enough defined over at least two decades. With this information,
        one could correct for the overestimated field strengths.

\end{enumerate}

\acknowledgments FH, EGZ and M-MML express special thanks to the
ITP at University of California at Santa Barbara (NSF Grant
no. PHY94-07194), the hospitality of which during the Astrophysical 
Turbulence Program has made possible this research. 
The work of EGZ was supported in part by NSF AST 98-00616.
M-MML was partially supported by an NSF CAREER Fellowship NSF AST 99-85392, and
by NASA ATP grant NAG5-10103. Computations
presented here were performed on SGI Origin 2000 machines of the
Rechenzentrum Garching of the Max-Planck-Gesellschaft and the National
Center for Supercomputing Applications (NCSA).
ZEUS-3D and ZEUS-MP were used by courtesy of the Laboratory for
Computational Astrophysics at the NCSA. We have made use of
NASA's Astrophysics Data System Abstract Service.

\cleardoublepage

%\begin{table}
%\begin{center}
%\caption{Model list}
%\begin{tabular}{ccccccc}
%\tableline\tableline
\begin{deluxetable}{ccccccc}
\tablecaption{Model list\label{tab:models}}
\tablehead{
    \colhead{Name}
   &\colhead{Resolution}
   &\colhead{$k_{drv}$}
   &\colhead{$\beta$}
   &\colhead{$M/M_{cr}$}
   &\colhead{$\lambda_J$}
   &\colhead{$n_J$}
}
\startdata
${\cal E}r1a$& $512^3$ & $1-2$& $4.04$ &$8.3$ &$0.501$ &$64$  \\
${\cal E}h1a$& $256^3$ & $1-2$& $4.04$ &$8.3$ &$0.501$ &$64$  \\
${\cal E}h1d$& $256^3$ & $1-2$& $0.20$ &$1.8$ &$0.501$ &$64$  \\
${\cal G}i1a$& $128^3$ & $1-2$& $4.04$ &$8.3$ &$1.067$ &$6.5$ \\
${\cal G}i1b$& $128^3$ & $1-2$& $1.13$ &$4.4$ &$1.067$ &$6.5$ \\
${\cal G}i1c$& $128^3$ & $1-2$& $0.50$ &$2.9$ &$1.067$ &$6.5$ \\
${\cal G}i1d$& $128^3$ & $1-2$& $0.18$ &$1.7$ &$1.067$ &$6.5$ \\
${\cal G}i1e$& $128^3$ & $1-2$& $0.13$ &$1.5$ &$1.067$ &$6.5$ \\
${\cal G}i1f$& $128^3$ & $1-2$& $0.09$ &$1.3$ &$1.067$ &$6.5$ \\
${\cal G}i1g$& $128^3$ & $1-2$& $0.07$ &$1.1$ &$1.067$ &$6.5$ \\
${\cal G}i1h$& $128^3$ & $1-2$& $0.05$ &$0.8$ &$1.067$ &$6.5$ \\
${\cal G}h1h$& $256^3$ & $1-2$& $0.05$ &$0.8$ &$1.067$ &$6.5$ \\
\tableline
$L$          & $256^3$ & $1-8$& $2.00$ &$\infty$&(...)&(...)\\
$MC81$       & $128^3$ & $7-8$& $2.00$ &$\infty$&(...)&(...)\\
$MA81$       & $128^3$ & $7-8$& $2.00$ &$\infty$&(...)&(...)\\
$MC41$       & $128^3$ & $3-4$& $2.00$ &$\infty$&(...)&(...)\\
\enddata
%\end{tabular}
\tablecomments{$\beta=P_{th}/P_{mag}=8\pi c_s^2 \rho/B^2$. 
$M/M_{cr}$ gives the ratio of cloud mass to critical mass according to 
equation~(\ref{equ:magstatsup}). $\lambda_J$ is the Jeans length, with 
a total box side length of $L=2$, and $n_J$ is the number of Jeans masses 
in the box. The non-selfgravitating models $L$, $MC81$, $MA81$ and $MC41$
are taken from Mac Low et al. (1998) and Mac Low (1999).}
\end{deluxetable}

\cleardoublepage

%\begin{table}
%\begin{center}
%\caption{Characteristic values of all models}
%\begin{tabular}{cccccc}
%\tableline\tableline
\begin{deluxetable}{cccccc}
\tablecaption{Characteristic parameters of all models\label{tab:turbcharacter}}
\tablehead{
              \colhead{Name}
             &\colhead{$\beta$}
             &\colhead{$\xi$}
             &\colhead{$E_{mag}/E_{kin}$}
             &\colhead{$\rmsB/\meanB^2$}
             &\colhead{$\sigma(\delta)\,[^o]$} 
}
\startdata
${\cal E}r1a$& $4.04$ &$0.96$ &$1.03$ &$13.06$ &$22.3$ \\
${\cal E}h1a$& $4.04$ &$0.35$ &$0.37$ &$12.00$ &$15.0$ \\
${\cal E}h1d$& $0.20$ &$0.52$ &$1.30$ &$ 0.79$ &$14.0$ \\
${\cal G}i1a$& $4.04$ &$0.21$ &$0.23$ &$ 8.90$ &$20.0$ \\
${\cal G}i1b$& $1.13$ &$0.36$ &$0.45$ &$ 3.90$ &$17.0$ \\
${\cal G}i1c$& $0.50$ &$0.69$ &$0.91$ &$ 3.10$ &$13.0$ \\
${\cal G}i1d$& $0.18$ &$0.48$ &$1.00$ &$ 0.88$ &$14.0$ \\
${\cal G}i1e$& $0.13$ &$0.31$ &$1.00$ &$ 0.45$ &$13.0$ \\
${\cal G}i1f$& $0.09$ &$0.28$ &$1.20$ &$ 0.32$ &$11.0$ \\
${\cal G}i1g$& $0.07$ &$0.27$ &$1.30$ &$ 0.27$ &$11.0$ \\
${\cal G}i1h$& $0.05$ &$0.24$ &$1.70$ &$ 0.17$ &$ 8.7$ \\
${\cal G}h1h$& $0.05$ &$0.26$ &$2.00$ &$ 0.15$ &$ 9.1$ \\
\tableline
$L1$         & $2.00$ &$0.83$ &$1.15$ &$2.55$ &$15.4$\\
$L2$         & $2.00$ &$1.07$ &$1.78$ &$1.53$ &$14.4$\\
$L3$         & $2.00$ &$1.03$ &$2.45$ &$0.73$ &$12.7$\\
$MC81$       & $2.00$ &$0.41$ &$0.50$ &$4.45$ &$16.4$\\
$MA81$       & $2.00$ &$0.79$ &$1.31$ &$1.51$ &$14.5$\\
$MC41$       & $2.00$ &$0.39$ &$0.48$ &$7.82$ &$15.4$\\
\enddata
%\end{tabular}
%\label{tab:turbcharacter}
\tablecomments{All values are taken at $t=0.0$ except for models $L1$ to $L3$, who are
a time series of decaying turbulence (Mac Low 1999). 
$\xi=E_{mag}^{turb}/E_{kin}$, the ratio of turbulent magnetic to turbulent
kinetic energy, where $E_{mag}^{turb}$ does not include the mean ($z$-)field contribution.
$E_{mag}/E_{kin}$ is the ratio of total magnetic energy (including
uniform component in $z$-direction) to kinetic energy, $B^2_{turb}/\meanB^2$ is the ratio 
between the turbulent magnetic energy ($=E_{mag}^{turb}$) and mean field energy, 
and $\sigma(\delta)$ gives the dispersion of field angle around the mean field direction 
in degrees.}
\end{deluxetable}

\cleardoublepage

%%%%%%%%%%%%%%%%%%%%%%%%%%%%%%%%%%%%
% polmap-part
%%%%%%%%%%%%%%%%%%%%%%%%%%%%%%%%%%%%

%%%%%%%%%%%%%%%%%%%%%%%%%%%%%%%%%%%%
% CF-part
%%%%%%%%%%%%%%%%%%%%%%%%%%%%%%%%%%%%
\clearpage

\end{document}